\pdfoutput=1
\documentclass[cernpreprint, atlasdraft=false, texlive=2023, UKenglish, texmf, orcidlogo]{atlasdoc}
 
\usepackage{atlaspackage}
\usepackage{atlasbiblatex}
 
\usepackage{atlasphysics}
\usepackage{atlasprocess}

\graphicspath{{logos/}{figures/}}
 
\usepackage{ANA-STDM-2022-06-PAPER-defs}
\usepackage{float}
\usepackage{longtable}

\AtlasTitle{Observation of electroweak production of $W^+W^-$ in association with jets in proton--proton collisions at $\sqrt{s}=13\,\mathrm{TeV}$ with the ATLAS Detector}

\AtlasAbstract{
A measurement of the production of $W$ bosons with opposite electric charges
in association with two jets is presented based on $140\,\mathrm{fb}^{-1}$
of data collected by the ATLAS detector in proton--proton collisions at
$\sqrt{s}=13\,\mathrm{TeV}$.
The analysis is sensitive to the scattering of $W$ bosons,
which is of particular interest in the ATLAS physics programme as it can be used
to probe the electroweak symmetry breaking mechanism of the Standard Model.
This signal is observed with a significance of 7.1 standard deviations above
the background expectation, while 6.2 standard deviations were expected.
The measured cross-section is determined in a signal-enriched fiducial
volume and is found to be $2.7\pm0.5\,\mathrm{fb}$, which is consistent with
the theoretical prediction of $2.20^{+0.14}_{-0.13}\,\mathrm{fb}$.
}

\AtlasRefCode{STDM-2022-06}
 
\PreprintIdNumber{CERN-EP-2024-015}

\AtlasJournal{JHEP}
\AtlasJournalRef{\JHEP 07 (2024) 254}
\AtlasDOI{10.1007/JHEP07(2024)254}

\hypersetup{pdftitle={ATLAS document},pdfauthor={The ATLAS Collaboration}}
 
\begin{document}
 
\maketitle
 
\tableofcontents

\section{Introduction}

% The next lines are included from the .//sections/introduction.tex input file
The scattering of $W$ bosons can be used to probe the electroweak (EWK)
symmetry breaking mechanism of the Standard Model (SM). Without a SM Higgs
boson, the scattering of longitudinally polarised $W$ bosons yields a
cross-section that violates unitarity at high centre-of-mass
energy~\cite{PhysRevD.16.1519}.
Should either the triple or quartic gauge couplings for these processes prove
to be inconsistent with the SM prediction, it will point towards the
necessity for a new description of the EWK sector of the
SM~\cite{PhysRevD.74.073005}.
 
\begin{figure}[!t]
\subfloat[]{
\includegraphics[width=0.283\textwidth]{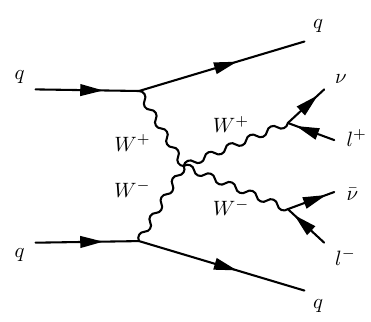}
\label{fig:Feynman:osWWewk1}}
\subfloat[]{
\includegraphics[width=0.362\textwidth]{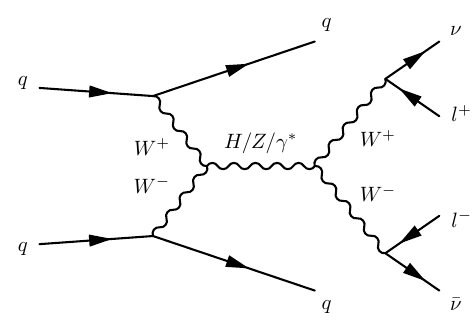}
\label{fig:Feynman:osWWewk2}}
\subfloat[]{
\includegraphics[width=0.326\textwidth]{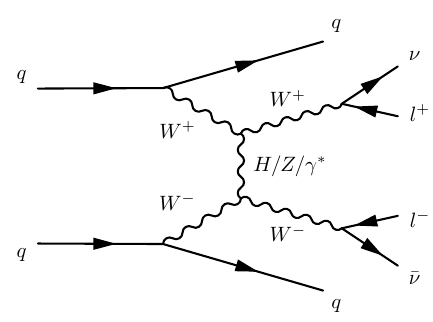}
\label{fig:Feynman:osWWewk3}}
\caption{\label{fig:Feynman:osWWewk} Examples of tree-level Feynman
diagrams for the electroweak production of \WWosjj events via the
vector boson scattering of two $W$ bosons radiated from the initial
partons.
The (a) quartic coupling of the $W$ bosons, (b) $s$-channel and (c)
$t$-channel diagrams are illustrated.
}
\end{figure}

This paper presents the observation by the ATLAS Collaboration
of the EWK production of \WWos in association with two jets (EWK \WWosjj),
which includes the scattering of $W$ bosons with opposite electric charges.
This
analysis is carried out with $140\,\mathrm{fb}^{-1}$ of proton--proton
collision data at $\sqrt{s}=13$~\TeV\ from LHC Run 2. The CMS Collaboration
previously observed this process using $138\,\mathrm{fb}^{-1}$ of
data~\cite{CMS:2022woe}.
The signal is characterised by one $W$ boson decaying into an electron and an
electron neutrino, and the other decaying into a muon and a muon neutrino.
This decay channel is chosen for its enhanced detection sensitivity over a
final state with leptons of the same flavour.
In this vector boson scattering (VBS) scenario,
the scattered $W$ bosons are radiated from the quarks that initiate the
hard interaction, so that the final state particles of the process at
leading order (LO) in the strong coupling constant
consist of two quarks that hadronise into jets, an electron, a muon,
and two neutrinos that manifest as missing transverse energy ($\met$) in
the ATLAS detector. Tree-level Feynman diagrams for the processes included in
the signal are shown in Figure~\ref{fig:Feynman:osWWewk}.
Figure~\ref{fig:Feynman:osWWewk1} illustrates the quartic coupling of
the $W$ bosons, while Figures~\ref{fig:Feynman:osWWewk2}
and~\ref{fig:Feynman:osWWewk3} represent the $s$-channel and $t$-channel VBS
diagrams, respectively.
 
\begin{figure}[!t]
\centering
\subfloat[]{
\includegraphics[width=0.320\textwidth]{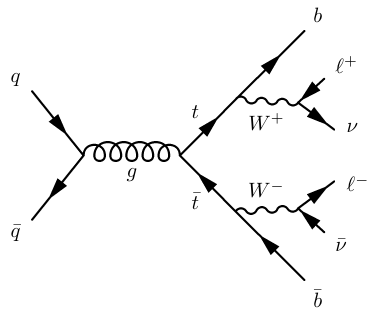}
}
\subfloat[]{
\includegraphics[width=0.320\textwidth]{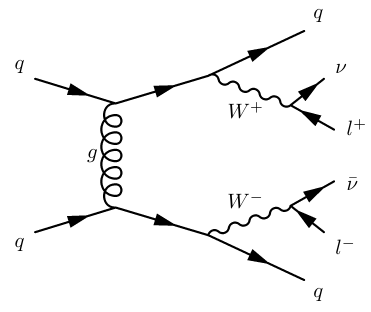}
}
\caption{\label{fig:Feynman:osWW_bgrd} Examples of tree-level Feynman
diagrams for the two main backgrounds: (a) \ttbar  and (b) strong
\WWosjj.
}
\end{figure}

A discriminant based on a neural network (NN) is employed to
distinguish the VBS signal from a large background dominated by
the production of top quarks (in pairs or singly).
This measurement is well suited to this approach due to the distinct kinematic
correlations between the physics objects in the
final state. In the VBS process, the radiation of the $W$ bosons from the
quarks
provides them with a boost that results in jets with high transverse momentum
($\pt$) entering the forward regions of the detector. The colourless exchange
between the $W$ bosons leads to a deficit of hadronic activity in the central
region of the detector compared to the background processes.
Figure~\ref{fig:Feynman:osWW_bgrd} displays the
diagrams for the two leading backgrounds, namely the production
of top quark pairs ($t\bar{t}$) and \WWosjj production by the strong
interaction (strong \WWosjj).
 
% End of text imported from the .//sections/introduction.tex input file

\section{ATLAS detector}

% The next lines are included from the .//sections/detector.tex input file
\newcommand{\AtlasCoordFootnote}{
ATLAS uses a right-handed coordinate system with its origin at the nominal interaction point (IP)
in the centre of the detector and the \(z\)-axis along the beam pipe.
The \(x\)-axis points from the IP to the centre of the LHC ring,
and the \(y\)-axis points upwards.
Polar coordinates \((r,\phi)\) are used in the transverse plane,
\(\phi\) being the azimuthal angle around the \(z\)-axis.
The pseudorapidity is defined in terms of the polar angle \(\theta\) as \(\eta = -\ln \tan(\theta/2)\) and is equal to the rapidity
$ y = \frac{1}{2} \ln \left( \frac{E + p_z }{E - p_z } \right) $ in the relativistic limit.
Angular distance is measured in units of \(\Delta R \equiv \sqrt{(\Delta y)^{2} + (\Delta\phi)^{2}}\).}

The ATLAS experiment~\cite{PERF-2007-01} at the LHC is a multipurpose particle detector
with a forward--backward symmetric cylindrical geometry and a near \(4\pi\) coverage in
solid angle.\footnote{\AtlasCoordFootnote}
It consists of an inner tracking detector surrounded by a thin superconducting solenoid
providing a \qty{2}{\tesla} axial magnetic field, electromagnetic and hadronic calorimeters, and a muon spectrometer.
The inner tracking detector covers the pseudorapidity range \(|\eta| < 2.5\).
It consists of silicon pixel, silicon microstrip, and transition radiation tracking detectors.
Lead/liquid-argon (LAr) sampling calorimeters provide electromagnetic (EM) energy measurements
with high granularity within the region \(|\eta|< 3.2\).
A steel/scintillator-tile hadronic calorimeter covers the central pseudorapidity range (\(|\eta| < 1.7\)).
The endcap and forward regions are instrumented with LAr calorimeters
for EM and hadronic energy measurements up to \(|\eta| = 4.9\).
The muon spectrometer surrounds the calorimeters and is based on
three large superconducting air-core toroidal magnets with eight coils each.
The field integral of the toroids ranges between \num{2.0} and \qty{6.0}{\tesla\metre}
across most of the detector.
The muon spectrometer includes a system of precision tracking chambers up to \(|\eta| = 2.7\) and fast detectors for triggering up to \(|\eta| = 2.4\).
The luminosity is measured mainly by the LUCID--2~\cite{LUCID2paper} detector, which is located close to the beampipe.
A two-level trigger system is used to select events~\cite{TRIG-2016-01}.
The first-level trigger is implemented in hardware and uses a subset of the detector information
to accept events at a rate below \qty{100}{\kHz}.
This is followed by a software-based trigger that
reduces the accepted event rate to \qty{1}{\kHz} on average
depending on the data-taking conditions.
A software suite~\cite{ATL-SOFT-PUB-2021-001} is used in data simulation, in the reconstruction
and analysis of real and simulated data, in detector operations, and in the trigger and data acquisition
systems of the experiment.

% End of text imported from the .//sections/detector.tex input file

\section{Data and simulated event samples}
\label{sec:samples}

% The next lines are included from the .//sections/samples.tex input file
The analysis is based on data collected by the ATLAS detector in proton--proton
collisions at a centre-of-mass energy of $13\,\TeV$ during \RunTwo of the LHC,
between 2015 and 2018.
After requiring the data to fulfil data quality criteria~\cite{DAPR-2018-01}, the
total luminosity of \RunTwo amounts to $140\,\mathrm{fb}^{-1}$, with an
uncertainty of 0.83\%~\cite{DAPR-2021-01}.

Monte Carlo (MC) predictions of SM processes were produced using
the ATLAS simulation infrastructure~\cite{SOFT-2010-01} that incorporates
the complete \GEANT~\cite{Agostinelli:2002hh} simulation of the ATLAS
detector. The effect of multiple interactions in the same or nearby bunch
crossing (\pileup) was simulated by generating a set of inelastic proton--proton
collision events using \PYTHIA[8.186]~\cite{Sjostrand:2007gs} 
with the A3 set of tuned parameters (tune)~\cite{ATL-PHYS-PUB-2016-017} and the
\nnpdflo~set 
of parton distribution functions
(PDF)~\cite{Ball:2012cx}.
These events are then overlaid on top of the original hard-scatter signal and
background events.
 
Signal \WWosjj events were modelled via a pure EWK prediction that
incorporates diagrams of order $\mathcal{O} (\alpha_\text{EWK}^{6})$
in perturbation theory. The Higgs boson contribution is suppressed by a
generator-level cut on the invariant mass of the decay leptons and neutrinos, while
$s$-channel contributions are neglected.
All the leptonic decays of the $W$ boson are considered,
with 11\% of the signal events selected by the analysis
involving at least one $W$ boson decaying into a $\tau$-lepton and a neutrino.
Events were simulated with \POWHEGBOX[v2]~\cite{Nason:2004rx,Frixione:2007vw,Alioli:2010xd,J_ger_2013} at next-to-leading order (\NLO) in quantum
chromodynamics (QCD) with the \nnpdf\ \NLO PDF
set~\cite{Ball:2014uwa}.
The sample was interfaced to \PYTHIA[8.244]~\cite{Sjostrand:2014zea} for the parton shower using
CTEQ6L1 PDF set~\cite{Pumplin:2002vw}
with the \AZNLO tune~\cite{STDM-2012-23} and the dipole recoil shower scheme,
preventing the generation of excess central jet radiation~\cite{ballestrero_precise_2018}.
 
The top-quark pair and single-top quark $Wt$ production were modelled using the
\POWHEGBOX[v2]~\cite{Frixione:2007nw}
generator at \NLO in QCD with the \nnpdf\ 
PDF set.
The events were interfaced to \PYTHIA[8.230] 
for the parton shower, hadronisation, and underlying event, with
parameters set according to the A14 tune~\cite{ATL-PHYS-PUB-2014-021} and
using the \nnpdflo\ 
PDF set. The mass of the top quark
was set to $m_\text{top} = 172.5$~\GeV.
The $h_\text{damp}$ parameter~\footnote{The resummation damping factor that
partly controls the matching of \POWHEG matrix elements to the parton shower
and thus effectively regulates the high-transverse-momentum radiation against
which the \ttbar system recoils.} was set to  1.5 $m_\text{top}$ for \ttbar events. For $Wt$
events, the four-flavour and five-flavour schemes were used to model the $t$-channel
and the $s$-channel, respectively.
The decays of bottom and charm hadrons were performed by
\EVTGEN[1.6.0]~\cite{Lange:2001uf}.
The \ttbar\ sample was normalised to the cross-section predicted at next-to-next-to-leading order (\NNLO) in
QCD including the resummation of next-to-next-to-leading-logarithmic (NNLL)
soft-gluon terms calculated using \textsc{Top++~2.0}~\cite{Beneke:2011mq,Cacciari:2011hy,Baernreuther:2012ws,Czakon:2012zr,Czakon:2012pz,Czakon:2013goa,Czakon:2011xx}, while the $Wt$ cross-section was corrected to the theory
prediction calculated at \NLO in QCD with NNLL soft-gluon corrections~\cite{Kidonakis:2010ux,Kidonakis:2013zqa}.

The \WWos production in association with two jets, originating from the strong interaction
$q\bar{q}\rightarrow W^+W^-$ and
$gg\rightarrow W^+W^-$ processes (strong \WWosjj), was
simulated with \SHERPA[2.2.2]~\cite{Bothmann:2019yzt}. The former relies on
matrix elements with up to one additional parton emission at \NLO accuracy
in QCD and up to three additional partons at \LO, while the latter allows up
to one additional parton at \LO and includes off-shell effects and Higgs
boson contributions. The virtual QCD correction was provided by the \OPENLOOPS
library~\cite{Cascioli:2011va,Denner:2016kdg}. The process initiated by
gluon-gluon fusion ($gg$) was
normalised to a cross-section at \NLO accuracy in QCD~\cite{Caola_2016}.
 
The strong production of $Z$/$\gamma^{*}$+jets events was simulated with
\SHERPA[2.2.1] 
using \NLO matrix elements for up to
two partons, and \LO matrix elements for up to four partons, as calculated with
the \comix~\cite{Gleisberg:2008fv} and
\OPENLOOPS 
libraries.
 
The strong production of $W$+jets was simulated with \MGNLO~\cite{alwall_automated_2014} with up to four partons at \LO in QCD\@. The sample was showered with
\PYTHIA[8.210] 
using the A14 tune 
and the \nnpdflo\ PDF set
and the cross-section was normalised to the \NNLO prediction from
\AtlasMC{FEWZ}~\cite{Anastasiou_2004}.
 
The $WZ$ and $ZZ$ processes were simulated with \POWHEGBOX[v2]
at \NLO in QCD\@. The samples were showered with \PYTHIA[8.210]
using the CTEQ6L1 PDF set
with the AZNLO tune. 
Only the strong production was considered, the weak production was found to be negligible.
Leptonic decays of WWW, WWZ and ZZZ productions were modelled with
\SHERPA[2.2.2]
using factorised gauge boson decays. Matrix elements were accurate at
\NLO in QCD for the inclusive process and at \LO for up to two additional
parton emissions. The virtual QCD correction for matrix elements was computed
at \NLO accuracy via the \OPENLOOPS
library.  
The $WZ$, $ZZ$, and triboson simulations are collectively referred to as `multiboson' production.
 
Higgs boson production from gluon--gluon fusion was simulated with
\POWHEG~NNLOPS~\cite{Hamilton:2013fea, Hamilton:2015nsa}.
The prediction was normalised to the N$^3$LO order cross-section in QCD and the \NLO cross-section in EWK~\cite{deFlorian:2016spz, Anastasiou:2016cez, Anastasiou:2015vya, Dulat:2018rbf, Harlander:2009mq, Harlander:2009bw, Harlander:2009my, Pak:2009dg, Actis:2008ug, Actis:2008ts, Bonetti:2018ukf}. The sample was showered with \PYTHIA[8.212]
using PDF4LHC15 PDF set~\cite{Butterworth:2015oua} with the AZNLO tune.
Higgs boson production from vector boson scattering was simulated with
\POWHEGBOX[v2]
at \NLO in QCD and interfaced to \PYTHIA[8.230] 
for parton showering.
Both samples described above are only used to tune the selection cuts, so
that such events have a negligible contribution to the events selected in the
analysis.
 
The matrix element calculations from all the samples generated with \SHERPA
were matched and merged with the parton shower from
\SHERPA~\cite{Schumann:2007mg} based on
Catani--Seymour dipole 
factorisation using the \MEPSatNLO
prescription~\cite{Hoeche:2011fd,Catani:2001cc,Hoeche:2009rj}.
The \nnpdf\ \NNLO 
PDF set was used along with a dedicated
set of tuned parton shower parameters developed by the authors of \SHERPA.

% End of text imported from the .//sections/samples.tex input file

\section{Event selection}

% The next lines are included from the .//sections/selection.tex input file
 
The candidate signal events are composed of $W$ pairs produced
in association with two or three jets, and decaying into an
electron and a muon with opposite electric charges plus neutrinos.
The analysis of this final state requires a proper reconstruction,
calibration and selection of these physics objects.

The events considered in the analysis were recorded by requiring a combination
of single electron and single muon triggers~\cite{TRIG-2018-05,TRIG-2018-01}.
The primary online electrons and muons considered are required to have a low
$\pt$ threshold ranging from $24\,\GeV$ to $26\,\GeV$ and from $20\,\GeV$ to
$26\,\GeV$, respectively.
Furthermore, these leptons must satisfy a loose to tight lepton quality
requirement, depending on the data taking period.
Additional triggers with looser lepton identification criteria, and with
$\pt$ thresholds from $60\,\GeV$ to $140\,\GeV$ for electrons and from
$50\,\GeV$ for muons, were used to increase the efficiency of the
event selection.
This exceeds 99\% for candidate events fulfilling the the selection criteria
described in Table~\ref{tag:selection:cuts}.

Electrons are reconstructed from energy deposits in the calorimeter that are
matched to tracks~\cite{EGAM-2021-01}. They are calibrated and required to
have $|\eta|=2.47$, excluding the transition region between the barrel and the
endcaps of the EM calorimeter, namely $1.37<|\eta|<1.52$.
Muons are reconstructed by combining the tracking information from the inner
detector and the muon spectrometer~\cite{MUON-2018-03}. They are calibrated
and required to have $|\eta|=2.5$.
Selected electrons and muons must fulfil their respective \texttt{Tight}
identification criteria, which relies on a likelihood in the case of electrons.

The primary vertex is selected from event candidates with the highest
$\sum\pt^2$ of the associated tracks recorded in the inner detector with
$\pt>500\,\MeV$, and both leptons are required to originate from it.
This requires the longitudinal impact parameter $z_0$ of the tracks associated
with the leptons to satisfy $|z_0\sin\theta|<0.5\,\mathrm{mm}$
and their transverse impact parameter significance $d_0/\sigma_{d_0}$ to fulfil
$|d_0/\sigma_{d_0}|<5$ and $<3$, for electrons and muons, respectively.
Leptons are further required to be isolated using information from the
tracks in the inner detector and from the energy clusters in the calorimeters
in a cone around the lepton.
The \texttt{Gradient}~\cite{EGAM-2018-01} criteria is used for electrons,
while muons are isolated according to the \texttt{Tight\_FixedRad} criteria,
which is close to the \texttt{Tight} working point described in
Reference~\cite{PERF-2015-10}, yet with improved background rejection for
muon $\pt>50\,\GeV$.
 
Jets are reconstructed using the anti-$k_{t}$ clustering
algorithm~\cite{Cacciari:2008gp,fastJetPaper}
with a radius parameter of $R=0.4$, and by using the particle flow
reconstruction~\cite{PERF-2015-09}, which combines calorimeter deposits and
tracks to determine jet properties.
They are calibrated to account for the detector response, including a
correction to their energy resolution~\cite{JETM-2018-05}.
Only jets within $|\eta|<4.5$ are
kept, and to suppress jets originating from additional proton--proton
interactions, those with $\pt<60\,\GeV$ and $|\eta|<2.4$ are required to
satisfy a \texttt{tight} jet-vertex tagger working point~\cite{PERF-2014-03}.
Jets with $\pt>20\,\GeV$ and $|\eta|<2.5$ originating from the decay of a
$b$-hadron are identified ($b$-tagged) using the \texttt{DL1r} $b$-tagging
algorithm~\cite{ATL-PHYS-PUB-2017-013,FTAG-2019-07} at the 85\% efficiency
working point.
 
The $\pt$ of the final state neutrinos can be inferred from the missing
transverse momentum $\vec{p}_T^{\,\mathrm{miss}}$ (with magnitude $\met$),
which is reconstructed from the calibrated leptons and jets.
The tracks from the inner detector that are not associated with a physics
object are also considered in the determination of
$\met$~\cite{ATLAS-CONF-2018-023}.

Candidate events are required to have a primary vertex associated with at
least two tracks, to ensure a proton--proton collision occurred.
Selected events are also required to contain at least one electron and one muon
with $\pt$ $>27\,\GeV$ and opposite electric charges.
Events with any additional electron satisfying the
\texttt{Loose}~\cite{EGAM-2018-01} 
isolation working point and likelihood-based
\texttt{Tight} 
(if $10\,\GeV<\pt<25\,\GeV$) or
\texttt{Medium}~\cite{EGAM-2021-01} 
(if $\pt>25\,\GeV$) criteria, are rejected.
Events containing an additional \texttt{Loose} 
muon with $\pt$ $>10\,\GeV$ satisfying the
\texttt{Loose}~\cite{MUON-2018-03} 
isolation working point are also rejected.
 
Only events containing either two or three jets with $\pt>25\,\GeV$ are
considered, as signal rates for higher multiplicities are significantly smaller.
The centrality of the leptons relative to the two leading jets in $\pt$
is defined as
\begin{equation}
\zeta = \mathrm{centrality} = \min \left\{ \left[ \min( \eta_{\ell_1}, \eta_{\ell_2} ) - \min( \eta_{j_1}, \eta_{j_2} ) \right],
\left[ \max( \eta_{j_1}, \eta_{j_2} ) - \max( \eta_{\ell_1}, \eta_{\ell_2} ) \right] \right\},
\end{equation}
 
where $\eta_{j_{1(2)}}$ and
$\eta_{\ell_{1(2)}}$ are the leading (subleading) jet and lepton pseudorapidities,
respectively.
To enhance the signal to background ratio the centrality is required to
exceed 0.5.
Due to the relative arrangement of the leptons and jets in an
EWK \WWosjj event, the signal process tends to produce events with a
positive lepton centrality.

Events surviving all these requirements constitute an inclusive sample in
which the final state objects are well measured and the kinematic features
that distinguish the signal from background processes are exploited to
improve the purity of the selected sample.
The production of Higgs boson mediated $W$ boson pairs via vector boson fusion is suppressed
by discarding events with an invariant mass of the two leptons ($m_{e\mu}$)
below $80\,\GeV$, and Drell--Yan events are reduced by requiring the $\met$ to
be above $15\,\GeV$ and by focusing on the $e\mu$ decay channel for the decay
of the $W$ boson pair.
The contribution of $W$ boson pairs with the
same electric charge is negligible given all the requirements above.
 
The definition of the signal region (SR) is completed by discarding
events containing at least one $b$-jet, which suppresses the
contribution from background processes involving top quark decays.
Although the SR is designed to specifically suppress certain backgrounds, a
small fraction of the EWK \WWosjj events survive the event selection.
The kinematic requirements on the different objects considered are summarised in
Table~\ref{tag:selection:cuts}, and the event yields for the different processes in the signal region, after the fit described
in Section~\ref{sec:result},
are provided in Table~\ref{tag:selection:yields}, distinguishing
between the two- and three-jet categories.
 
Finally, the discrimination between signal and background in the SR is
performed by a NN, defined in Section~\ref{sec:nn}, which is used to
extract the signal via a likelihood fit.

\begin{table}[!t]
\centering
\caption{ Selection cuts on physics objects and events that define the
signal region. }
\begin{tabular}{ll}
\toprule
Category & Requirements\\
\midrule
Leptons  & $\pt>27\,\GeV$ \\
& $|\eta|<2.47$ excluding $1.37<|\eta|<1.52$ (electrons) \\
&  $|\eta|<2.5$ (muons) \smallskip \\
& Identification: \texttt{Tight} \\% (electrons),
& Isolation: \texttt{Gradient} (electrons),
\texttt{Tight\_FixedRad} (muons) \smallskip \\
& $|d_0/\sigma_{d_0}|<5$ (electrons), $|d_0/\sigma_{d_0}|<3$ (muons) \\
& $|z_0\sin\theta|<0.5\,\mathrm{mm}$ \smallskip \\
 
\smallskip
$b$-jets & $\pt>20\,\GeV$ and $|\eta|<2.5$ $\;$
(\texttt{DL1r}
$b$-tagging with 85\% efficiency)
\\
Jets     & $\pt>25\,\GeV$ and $|\eta|<4.5$ \\
\midrule
Events   & One electron and one muon with opposite electric charges \\
 
& No additional lepton with $\pt>10\,\GeV$,
\texttt{Loose} isolation, \\
& \texttt{Tight}/\texttt{Medium} (electrons) and \texttt{Loose} (muons) identification
\\
 
& $m_{e\mu}>80\,\GeV$ \\
& $\met>15\,\GeV$ \\
& No $b$-jet \\
& Two or three jets \\
& $\zeta>0.5$ \\
 
\bottomrule
\end{tabular}
\medskip
\label{tag:selection:cuts}
\end{table}

\begin{table}[!t]
\centering
\caption{ The composition of the events in the signal region predicted by the
SM and the total predicted yield compared with the data. The events are split
into two categories, depending on the number of jets.
The uncertainties include both statistical and systematic contributions,
and correspond to the values after the likelihood fit described in
Section~\ref{sec:result}.
The uncertainties in the total predictions are smaller than the individual
component uncertainties due to correlations induced by the fit.
}
\begin{tabular}{
l
r@{$\:\pm\:$}l
r@{$\:\pm\:$}l
}
\toprule
& \multicolumn{4}{c}{Event yields} \\
Process        & \multicolumn{2}{c}{$n_{\text{jets}}=2$} & \multicolumn{2}{c}{$n_{\text{jets}}=3$} \\
\midrule
EWK \WWosjj    &  $158$ & $27$  & $54$   & $13$ \\
\ttbar         & $2394$ & $194$ & $1625$ & $125$ \\
Single top     & $491$  & $34$  & $225$  & $21$ \\
Strong \WWosjj & $1214$ & $256$ & $514$  & $121$ \\
\Wjets         & $37$   & $97$  & $19$   & $48$ \\
\Zjets         & $216$  & $62$  & $65$   & $25$ \\
Multiboson     & $101$  & $5$   & $42$   & $3$ \\
\midrule
SM prediction  & $4610$ & $77$ & $2546$ & $48$ \\
Data           & \multicolumn{2}{c}{$4610$} & \multicolumn{2}{c}{$2533$} \\
\bottomrule
\end{tabular}
\medskip
\label{tag:selection:yields}
\end{table}

% End of text imported from the .//sections/selection.tex input file

\section{Neural network discriminant}
\label{sec:nn}

% The next lines are included from the .//sections/neuralnet.tex input file
A neural network is used to separate between the signal and the main
remaining backgrounds from the SM, which constitute a large fraction of the SR.
The discrimination power of the NN is improved by training it
separately for SR events including exactly two or three jets, which
differ by the radiation of an additional gluon in the case of signal.
The training and evaluation of the NN are achieved with the
Keras~\cite{chollet2015keras} library interfaced within TMVA~\cite{TMVA},
by only
considering the dominant top quark and strong \WWosjj processes as backgrounds.
The signal and the background samples are each split into a training and a
testing components of equal sizes.
Two hidden layers with 108 nodes on the first hidden layer and 60 on the
second one are implemented.
The output layer incorporates two neurons and the final output of the network
takes values between 0 and 1.
The optimisation of the hyperparameters,
including the number of neurons and layers,
is accomplished by maximizing
the resulting area under the ROC
curve,\footnote{Receiver Operating Characteristic curve.} via a grid search.
This figure of merit also influences the choice of the input variables to the NN, which
are mostly weakly correlated between each other.

The set of input variables used in the two-jet category consists of
the leading and subleading jet $\pt$, $m_{\ell\ell}$, $\zeta$, the $\met$
significance,\footnote{$\met$ significance is computed as $\frac{|\vec{p}_T^{\,\mathrm{miss}}|}{ \sqrt{\sigma_\text{L}^{2}(1-\rho_\text{LT}^{2})}}$, where $\sigma_\text{L}$ is the total variance in the direction longitudinal to $\met$, and $\rho_\text{LT}$ is the correlation coefficient of the longitudinal (L) and transverse (T) $\met$
measurements.}
the difference $\Delta\eta_{jj}$ between the pseudorapidities of the two
leading jets, the azimuthal angle separation $\Delta\phi_{jj}$ between the
two leading jets, and the invariant mass $m_{e\mu jj}$ constructed
from the four-vectors of the two leading leptons and the two leading jets.
In the case of the three-jet category, the same input variables are used with
the addition of the $\pt$ and the centrality of the third-leading jet,
the latter being
defined as 
 
\begin{equation}
\zeta_3 = \mathrm{third\text{-}leading\;jet\;centrality} = \left| y_{j_3} - \frac{1}{2}\cdot\frac{y_{j_1} + y_{j_2}}{y_{j_1} - y_{j_2}} \right|,
\end{equation}
 
where $y_{j1}$, $y_{j2}$ and
$y_{j3}$ are the rapidities of the leading, subleading and third-leading jets
in $\pt$, respectively.
For both jet categories, the most sensitive variables in the NN are the $\pt$
of the two leading jets, as they tend to have larger values in VBS processes.
Furthermore, in the three-jet category, the third-leading jet has a centrality
($\zeta_3$) that peaks at high values in signal events, as it is
frequently emitted from one of the two leading forward jets, thus providing
an additional gain in sensitivity.
The invariant mass $m_{jj}$, constructed from the two leading jets in $\pt$,
although usually considered to enhance a VBS phase space, is not used in the
input to the NN because it can largely be derived from the observables
already used.
 
To avoid overtraining the NN and to increase its robustness, a dropout
regularisation with a rate of 0.1 is applied to both hidden layers.
This procedure randomly drops some neurons from the NN at each step of the
training, which can be interpreted as considering the average of different
layer set-ups. The evaluation of the NN using either the training or an
independent test sample produces compatible responses. The NN is further
validated in the SR by comparing data and simulation for the correlations of the
input variables between each other and against the NN output.
Finally, the binning of the NN output is optimised to increase the sensitivity
to the signal at high NN output values.

% End of text imported from the .//sections/neuralnet.tex input file

\section{Background determination}

% The next lines are included from the .//sections/background.tex input file
The main background to the EWK production of \WWosjj events consists of top
quarks produced either individually or in pairs, and in total represents
66\% of the SR\@.
The second most important background is the strong production of pairs of
$W$ bosons in association with jets, which amounts to 24\% of the
SR\@.
The remaining backgrounds are the production of a $Z$ boson in association with
jets (4\%), other multiboson events (2\%), and a $W$ boson in
association with jets where one jet is misidentified as a lepton (below 1\%).

\begin{figure}[!t]
\centering
\includegraphics[width=0.45\textwidth]{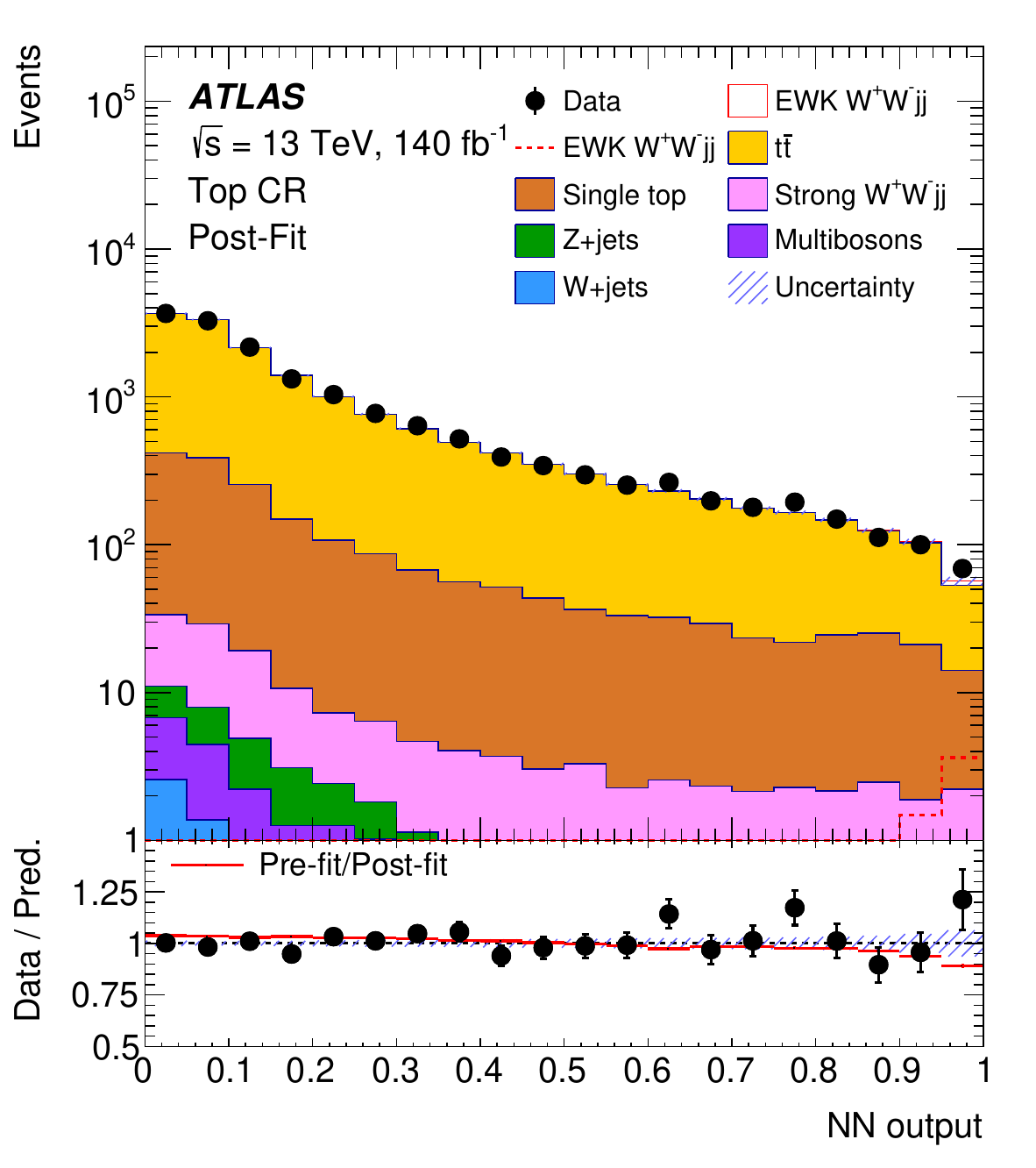}
\caption{ Distribution of the NN output in the top quark CR, with the top quark, strong \WWosjj, and
signal processes constrained by the likelihood fit
described in Section~\ref{sec:result}.
The two- and three-jet categories are merged.
Data corresponds to the black dots, the signal normalised to its
simulated cross-section is represented by the unstacked dashed
histogram, and the post-fit predictions of the SM together with the
signal are depicted by the stacked filled histograms.
The lower panel shows the ratio of the data to the post-fit
predictions of the SM, and the ratio of the
pre-fit to the post-fit SM yields.
The uncertainty in the total contribution from the SM is
illustrated by a hashed band.
}
\label{fig:postfit_nn_cr}
\end{figure}

The background originating from the production of top quarks is modelled by
simulation, and further constrained with data in a dedicated control
region (CR). The latter is defined with the same cuts as the SR except for
requiring one of the two leading jets to be a $b$-jet.
This region is thus dominated by events including a top quark,
as highlighted in Figure~\ref{fig:postfit_nn_cr}, which shows the NN
discriminant evaluated from events in the CR, using the same model as in
the SR for the two- and three-jet categories.
In the NN distribution, signal-like and background-like events tend to have
NN output values closer to 1 and 0, respectively.
The normalisation of the predicted top quark production is constrained by a
binned likelihood fit that includes both the CR and the SR,
as described in Section~\ref{sec:result}.
The top quark CR proves to be
necessary to constrain the normalisation of top quark events, since the low
bins of the NN output in the SR, although enriched in top
quarks, are found not to have a sufficient constraining power due to the substantial number of strong \WWosjj events
with similar NN shape in the SR.

It is challenging to define a dedicated CR to enhance the strong production of
\WWosjj events, given the significant contribution of events originating from
the production of top quarks.
Nevertheless, the normalisation of simulated \WWosjj events is left as a free
parameter in the likelihood fit 
in order to constrain it with data together with the top quark background,
using events from the top quark CR and SR.

The less relevant background consisting of $W$ boson production in association
with jets is modelled by simulation. To better account for this fraction of
events where jets are misreconstructed as leptons, the 
prediction is scaled up by 15\% to 60\% across the NN distribution using a
constraint from a data region enriched in fake leptons.
The latter is defined by selecting leptons with a looser identification, and
by requiring that at least one of the two leptons fails to satisfy the isolation and the
tight identification requirements, the other selection cuts being the same as
in the SR.
The semileptonic decays of top quarks is accounted for in the fake region as
it includes fake leptons coming from $b$-jets.

Since only a few events arising from the $Z$ boson and multiboson predictions
survive the SR selection, these backgrounds are modelled by
simulation with no additional constraints from data.
 
\begin{figure}[!t]
\centering
\subfloat[]{
\includegraphics[width=0.45\textwidth]{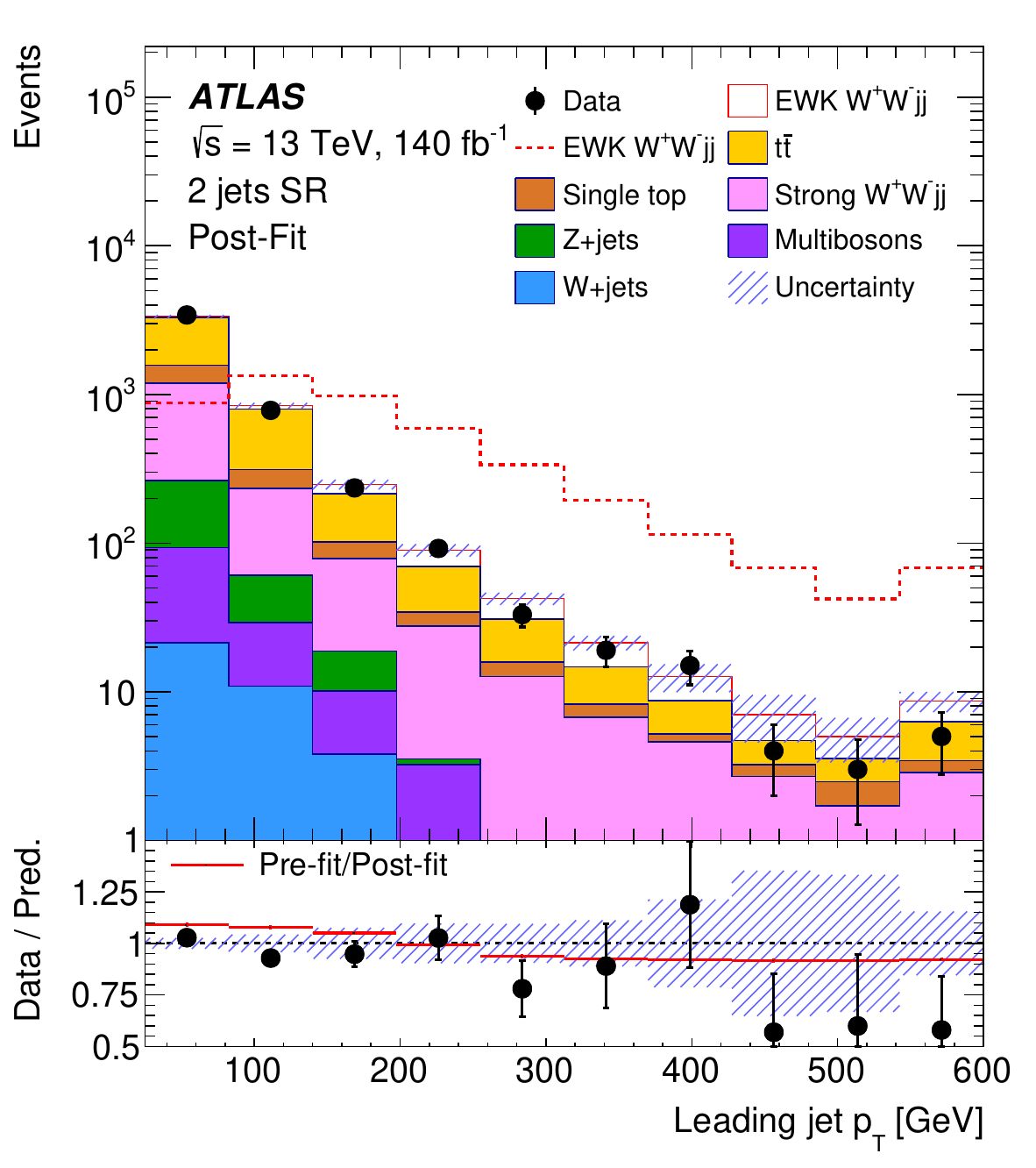}
}
\subfloat[]{
\includegraphics[width=0.45\textwidth]{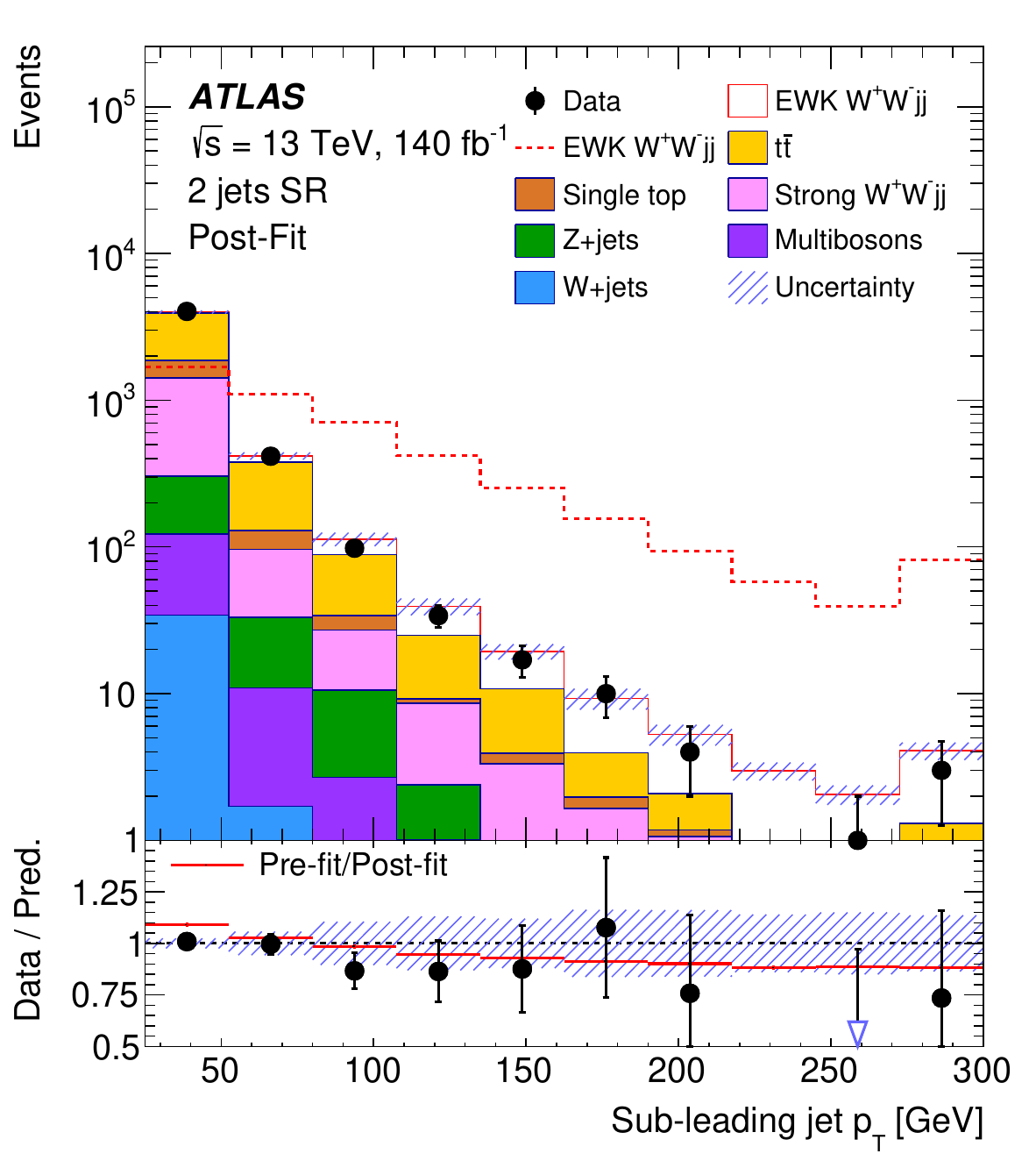}
}\\
\subfloat[]{
\includegraphics[width=0.45\textwidth]{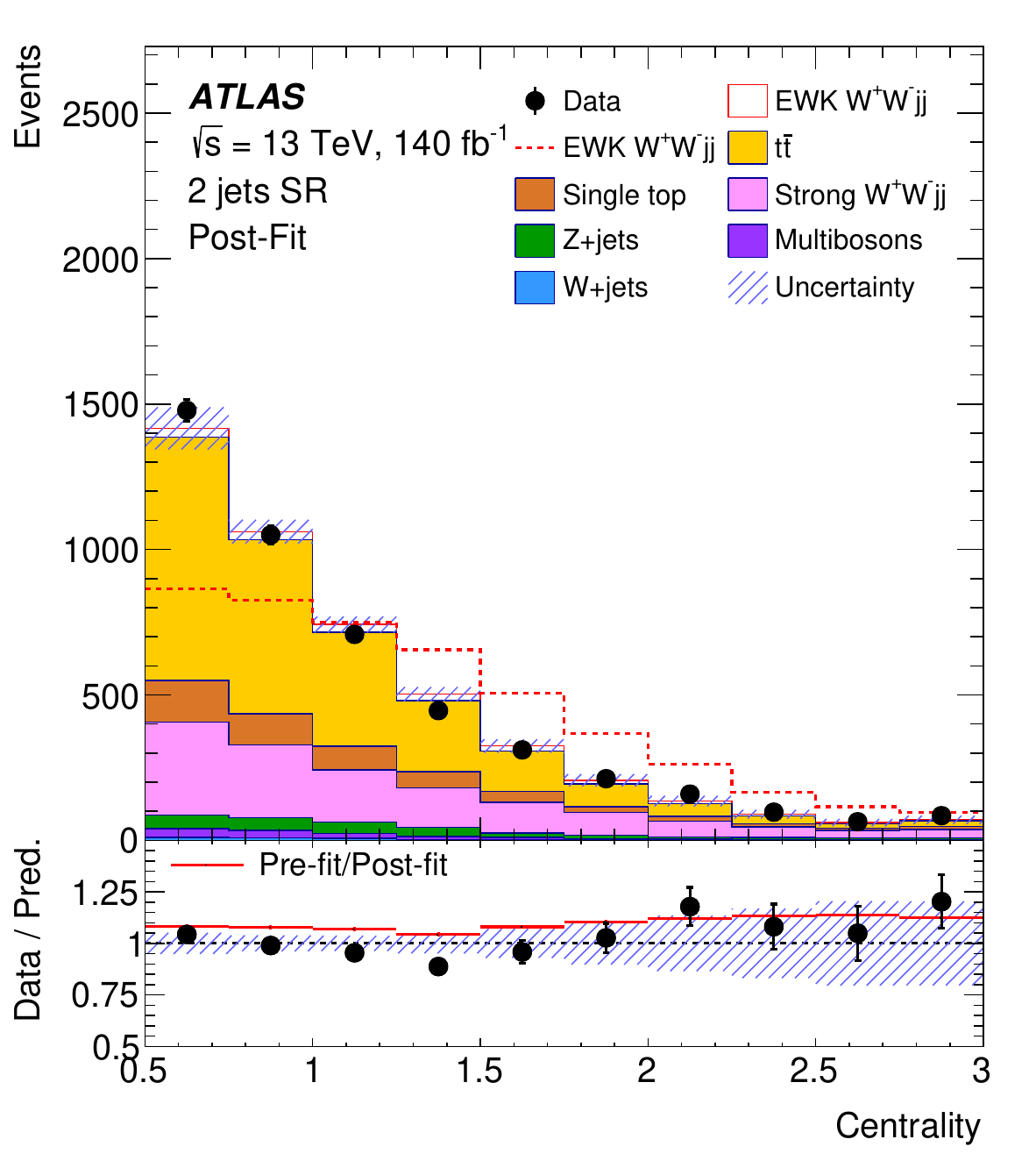}
}
\subfloat[]{
\includegraphics[width=0.45\textwidth]{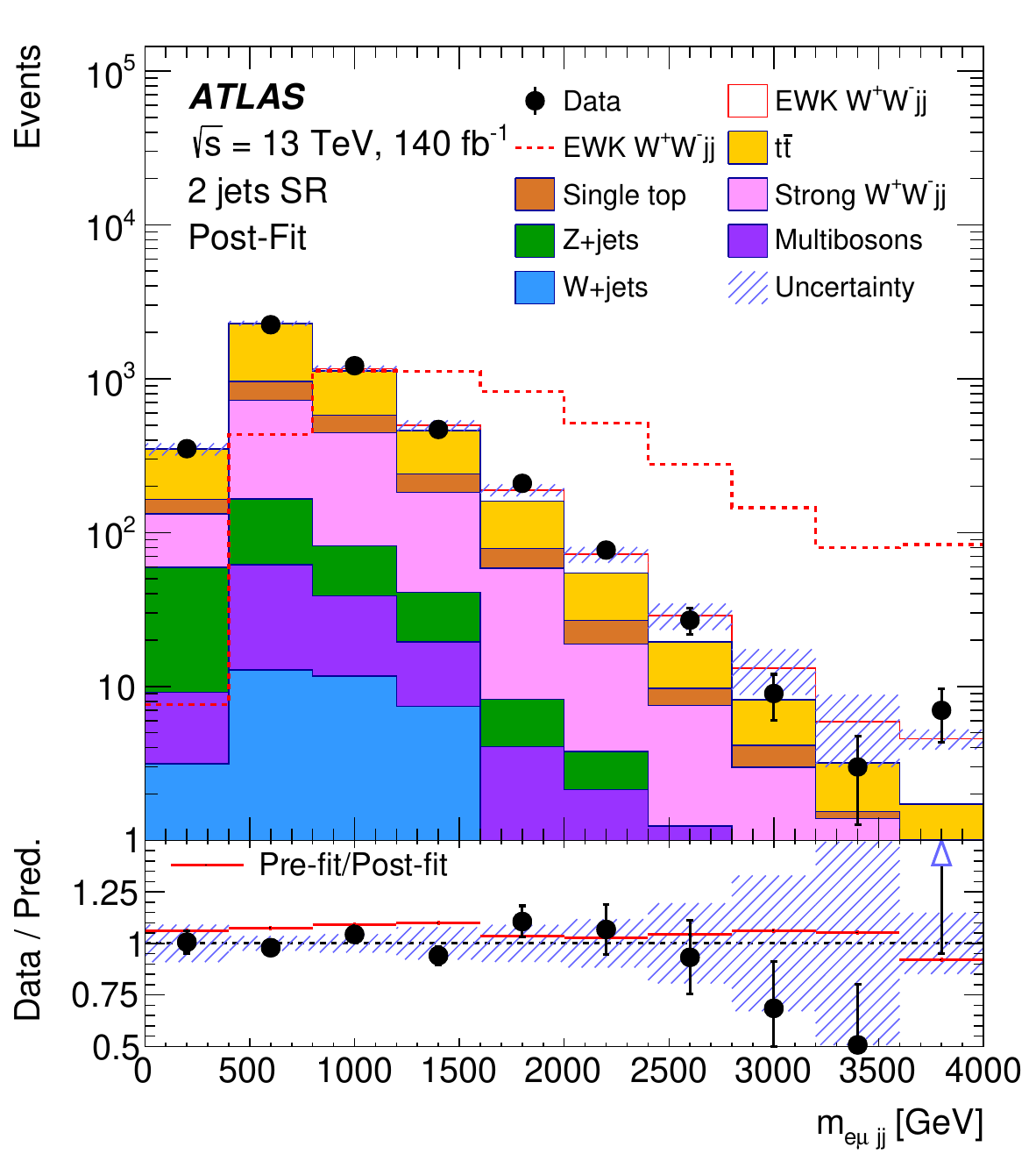}
}\\
\caption{ Distributions of (a) the leading  and (b) subleading  jet $\pt$,
(c) the centrality  and (d) the invariant mass $m_{e\mu jj}$
constructed from the four-vectors of the two leading leptons and
jets.
The observables are presented in the SR in
the two jet case, with the top quark, strong \WWosjj,
and signal processes constrained by the likelihood fit described
in Section~\ref{sec:result}.
Data corresponds to the filled dots, the signal normalised to the
total SM background is represented by the unstacked dashed
histogram,
and the post-fit predictions of the SM together with the signal
are depicted by the stacked filled histograms.
The lower panels show the ratios of the data to the post-fit
predictions of the SM, and the ratio of the
pre-fit to the post-fit SM yields.
The uncertainty in the total contribution from the SM is
illustrated by a hashed band.
}
\label{fig:sig_region_2jets}
\end{figure}

\begin{figure}[!t]
\centering
\subfloat[]{
\includegraphics[width=0.45\textwidth]{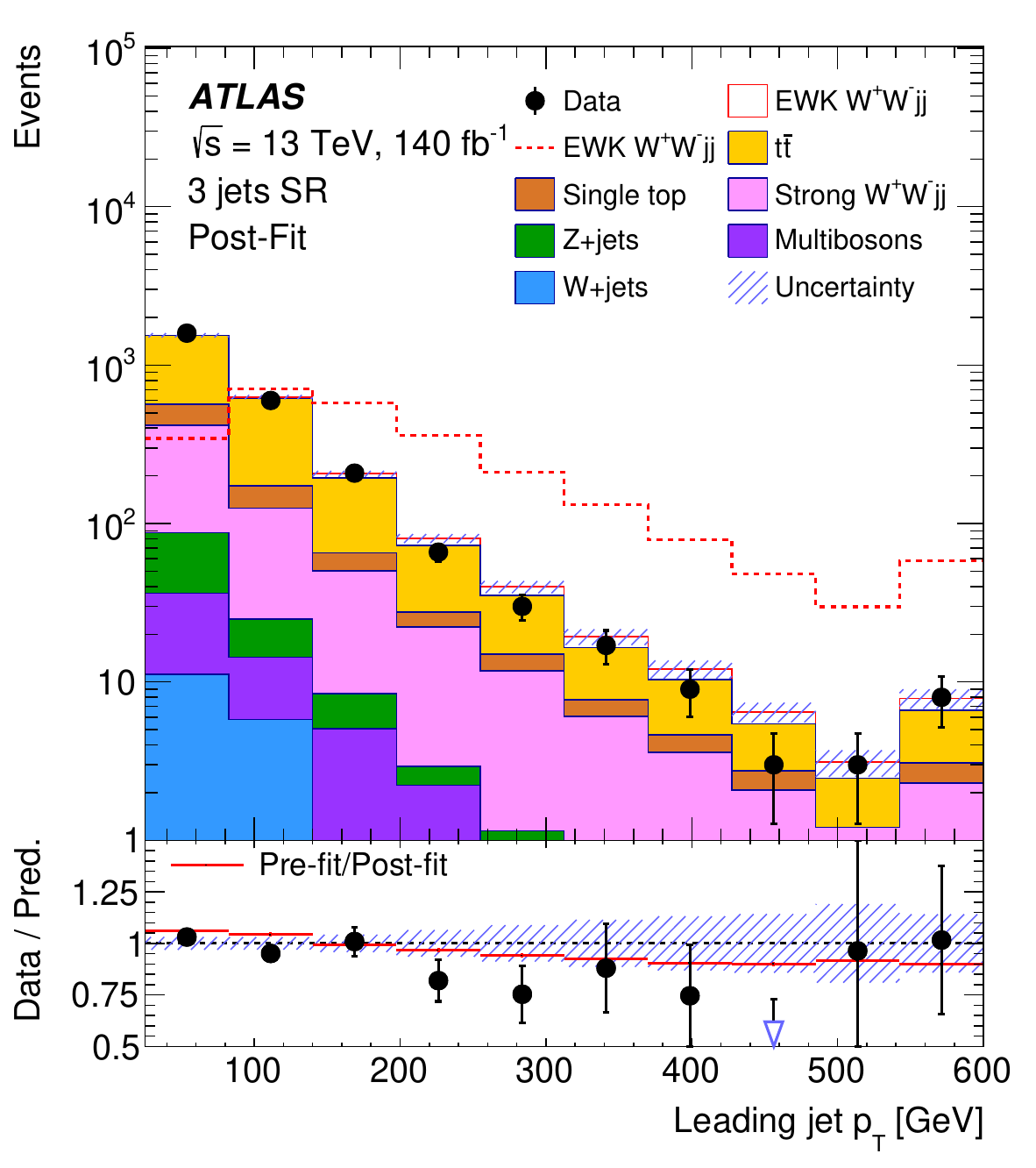}
}
\subfloat[]{
\includegraphics[width=0.45\textwidth]{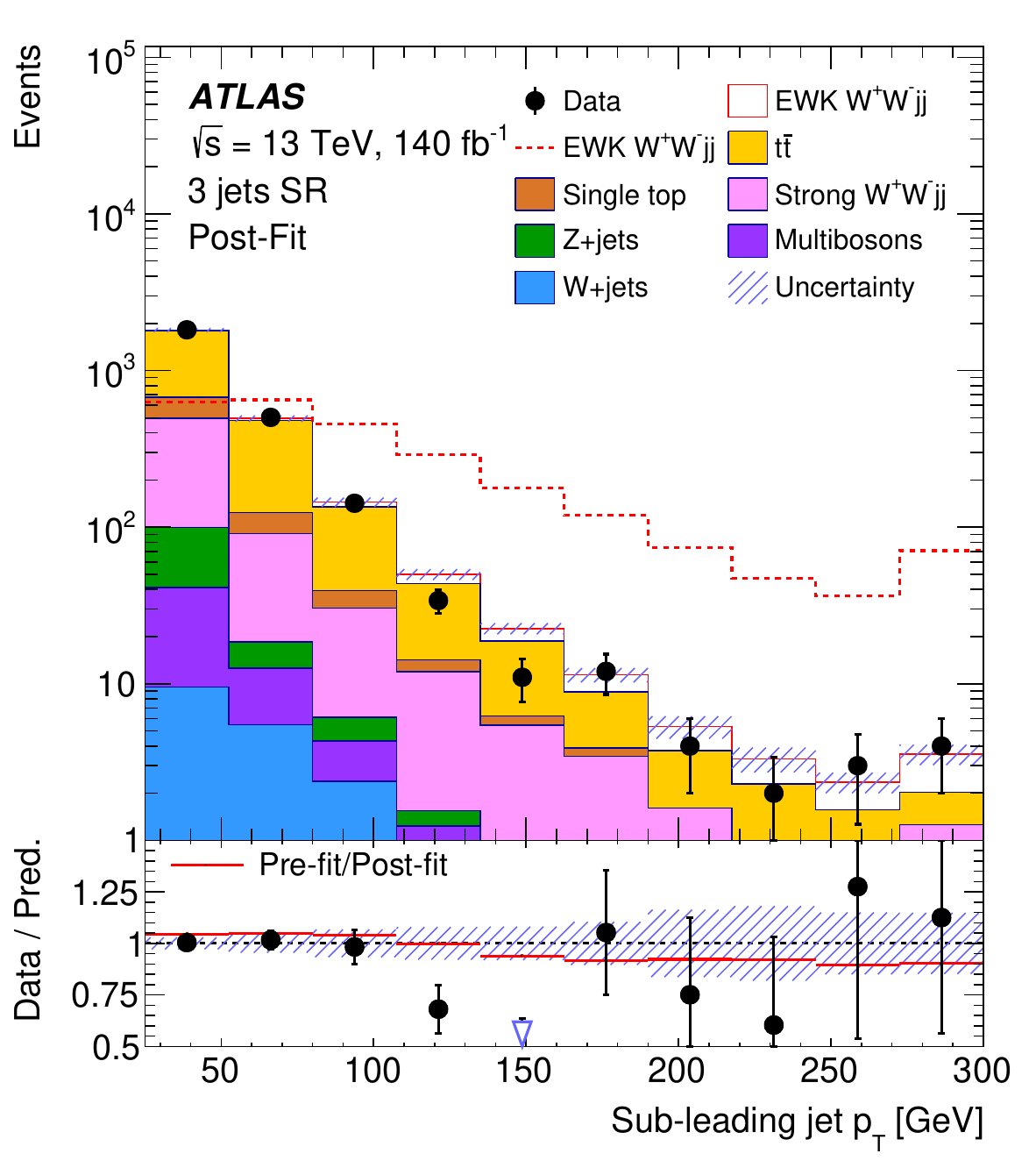}
}\\
\subfloat[]{
\includegraphics[width=0.45\textwidth]{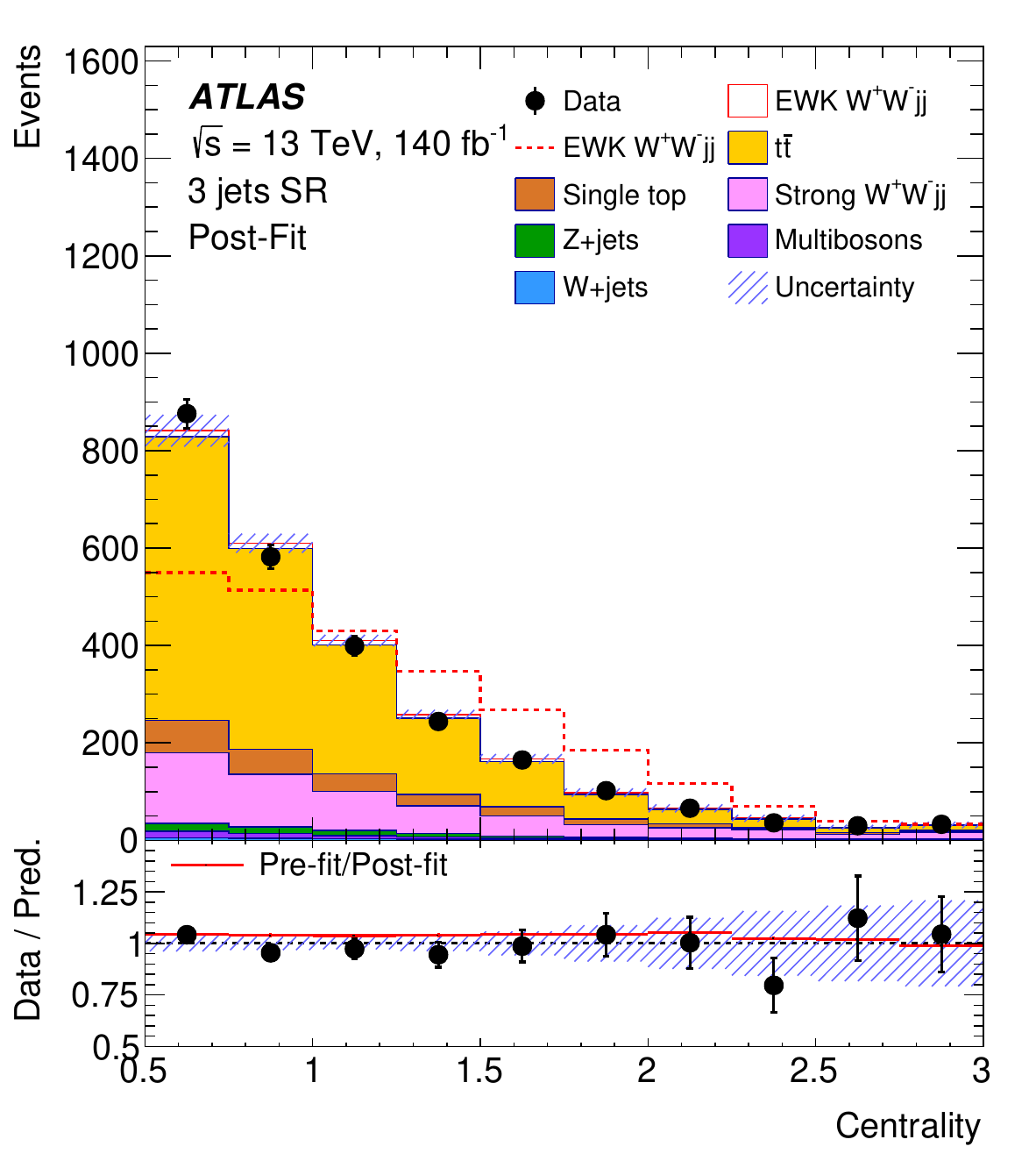}
}
\subfloat[]{
\includegraphics[width=0.45\textwidth]{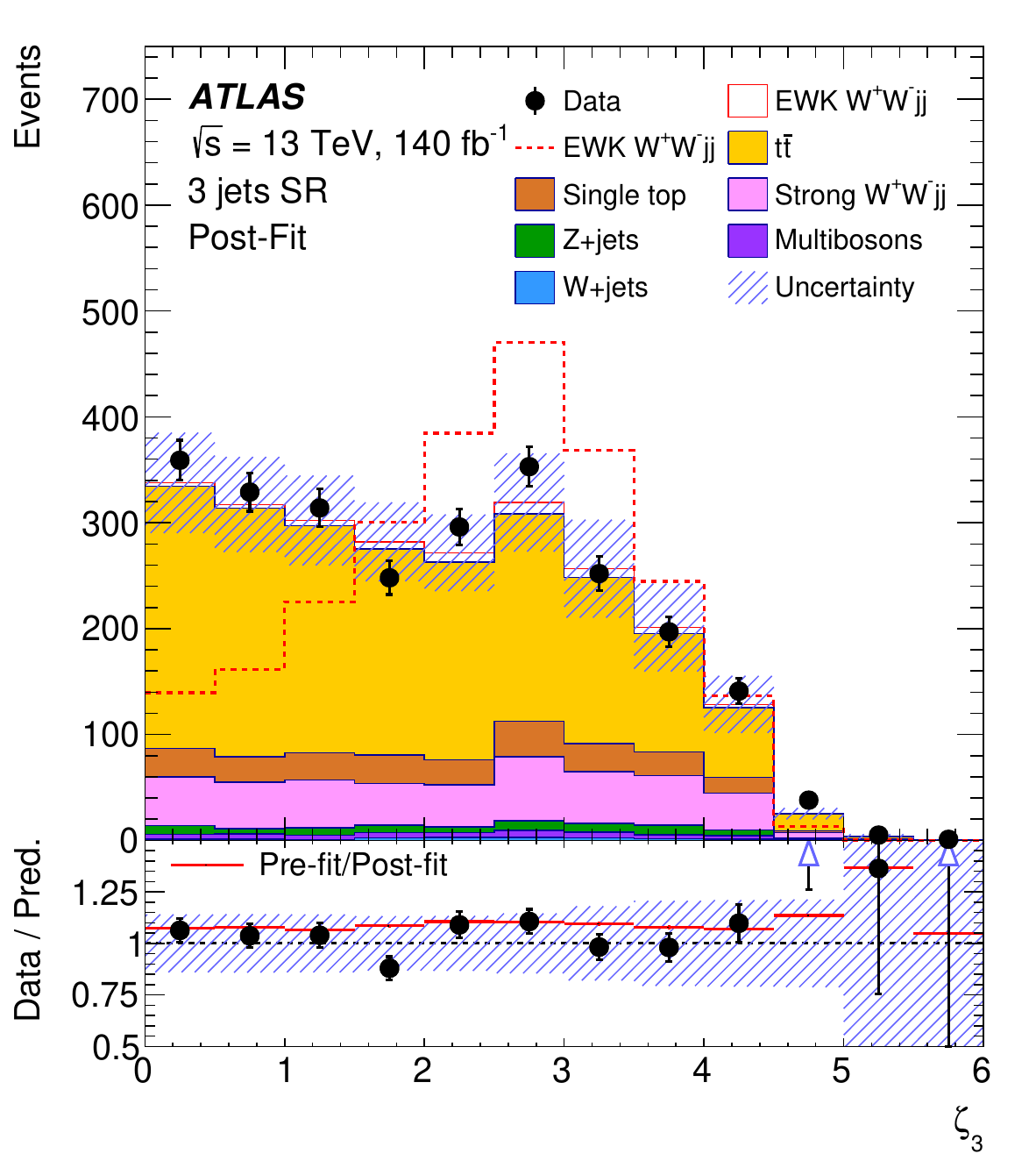}
}\\
\caption{ Distributions of (a) the leading  and (b) subleading jet $\pt$,
(c) the centrality  and (d) the third-leading jet centrality
$\zeta_3$.
The observables are presented in the SR in the three jet case, with the top quark, strong \WWosjj,
and signal processes constrained by the likelihood fit described
in Section~\ref{sec:result}.
Data corresponds to the filled dots, the signal normalised to the
total
SM background is represented by the unstacked dashed histogram,
and the post-fit predictions of the SM together with the signal
are depicted by the stacked filled histograms.
The lower panels show the ratios of the data to the post-fit
predictions of the SM, and the ratio of the
pre-fit to the post-fit SM yields.
The uncertainty in the total contribution from the SM is
illustrated by a hashed band.
}
\label{fig:sig_region_3jets}
\end{figure}

Figure~\ref{fig:sig_region_2jets} shows the distributions of the leading and
subleading jet $\pt$, the centrality, and the invariant mass $m_{e\mu jj}$
constructed from the four-vectors of the selected leptons and jets in the
two-jet signal region.
Figure~\ref{fig:sig_region_3jets} shows the distributions of the leading
and subleading jet $\pt$, the centrality, and the third-leading jet centrality
in the three-jet signal region. All figures are shown after the binned
likelihood fit described in Section~\ref{sec:result}, and demonstrate a good
overall modelling of these observables, which are used as input to the NN\@.
However, some variables like the leading jet $\pt$ are sensitive to the
mismodelling of the top quark $\pt$, which is not corrected for. The potential
subsequent tensions in the agreement between data and the SM prediction are
however mostly covered at the edge of the uncertainty bands.
 
% End of text imported from the .//sections/background.tex input file

\section{Uncertainties}

% The next lines are included from the .//sections/uncertainties.tex input file
Systematic uncertainties in the measurement of the signal cross-section originate from experimental and theoretical sources due to the signal and background modelling.
 
The dominant experimental uncertainties are related to the calibration of the jets, namely the jet energy scale and resolution~\cite{JETM-2018-05}, the $b$-tagging efficiency 
and the jet flavour composition.
Other sources of experimental uncertainties are due to the calibration of the leptons, which affects their identification and isolation efficiencies,
their energy resolution and momentum scale~\cite{EGAM-2021-01,MUON-2022-01},
and the lepton trigger efficiencies 
and pile-up modelling.
To account for the mismodelling of the simulated $W$+jets background where one
jet is misidentified as a lepton, a conservative uncertainty of 100\% is
assigned to its normalisation, with a negligible impact on the signal
extraction.
As this uncertainty is modelled by a gaussian in the likelihood fit, the
corresponding postfit errors provided in Table 2 overlaps with negative
estimates.
All experimental uncertainties are propagated through the analysis and the final impact on the signal strength is evaluated when performing the likelihood fit.

The uncertainties in the modelling of the signal includes variations of  the renormalisation and factorisation scales, the PDF, the parton shower and the initial- and final-state radiation.
To estimate the uncertainty due to missing higher-order QCD corrections, the renormalisation and factorisation scales are varied up and down by factors of two, avoiding opposite variations.
The parton shower uncertainty is estimated by comparing the nominal simulation, described in Section~\ref{sec:samples}, with an alternative prediction performed using \POWHEGBOX[v2] 
interfaced to \HERWIG[7.2.1]~\cite{Bahr:2008pv,Bellm:2015jjp} with the dipole shower model and the H7.1-Default hadronisation and underlying-event tune. The interference between EWK and strong production of \WWosjj was evaluated using \MGNLO, 
and the resulting value was taken as an uncertainty in the signal, amounting to 1\%.
 
Theoretical uncertainties in the prediction of the background are mostly due to  the simulation of top quark events. The parton shower and hadronisation modelling is evaluated by comparing the nominal \POWPY[8] prediction with an alternative one performed using \POWHEGBOX[v2] interfaced to \HERWIG[7.04]
with the H7UE set of tuned parameters 
and the \textsc{MMHT2014} LO PDF set~\cite{Harland-Lang:2014zoa}. Similarly, the uncertainty in the matching of NLO matrix elements to the parton shower is evaluated by comparing the nominal simulation with events generated by \MGNLO[2.6.2]
at NLO in QCD with the five-flavour scheme and the \textsc{NNPDF2.3} NLO PDF set and interfaced to \PYTHIA[8.230]. 
The uncertainty due to higher-order QCD effects and initial-state radiation for the top quark background is evaluated by varying the renormalisation and factorisation scales, the $h_\text{damp}$ parameter, and the VAR3c up and down variants of the A14 tune~\cite{ATL-PHYS-PUB-2017-007}. The final-state radiation uncertainty is evaluated by scaling up and down by a factor of two the renormalisation scale used for final-state parton shower emissions.
The uncertainty associated with the $Wt$ interference with \ttbar is estimated by comparing the nominal $Wt$ simulation, where \ttbar contributions are removed at the amplitude level using the diagram-removal scheme~\cite{Frixione:2008yi}, with an alternative $Wt$ simulation, where \ttbar contributions are removed at the cross-section level using the diagram-subtraction scheme~\cite{Frixione:2008yi}.
 
Theoretical uncertainties in the strong $qq \rightarrow WW$ and
$Z+\mathrm{jets}$
backgrounds originate from the renormalisation and factorisation scales,
and are estimated by using the same method as for the signal.
The theoretical uncertainty on the $gg \rightarrow WW$ production is neglected as this background represents
less than 7\% of the $qq \rightarrow WW$ process, and has no impact on the final result.
 
Finally, PDF uncertainties are estimated by using a reweighting procedure at the
matrix element level for 100 variations of the \nnpdf\ \NLO
PDF set for the signal and for the top quark, strong \WWosjj, and
$Z+\mathrm{jets}$ backgrounds.
For each bin of the NN, the standard deviation of these variations is taken as the uncertainty.

% End of text imported from the .//sections/uncertainties.tex input file

\section{Signal extraction and cross-section measurement}
\label{sec:result}

\begin{figure}[!t]
\centering
 
\subfloat[]{
\includegraphics[width=0.45\textwidth]{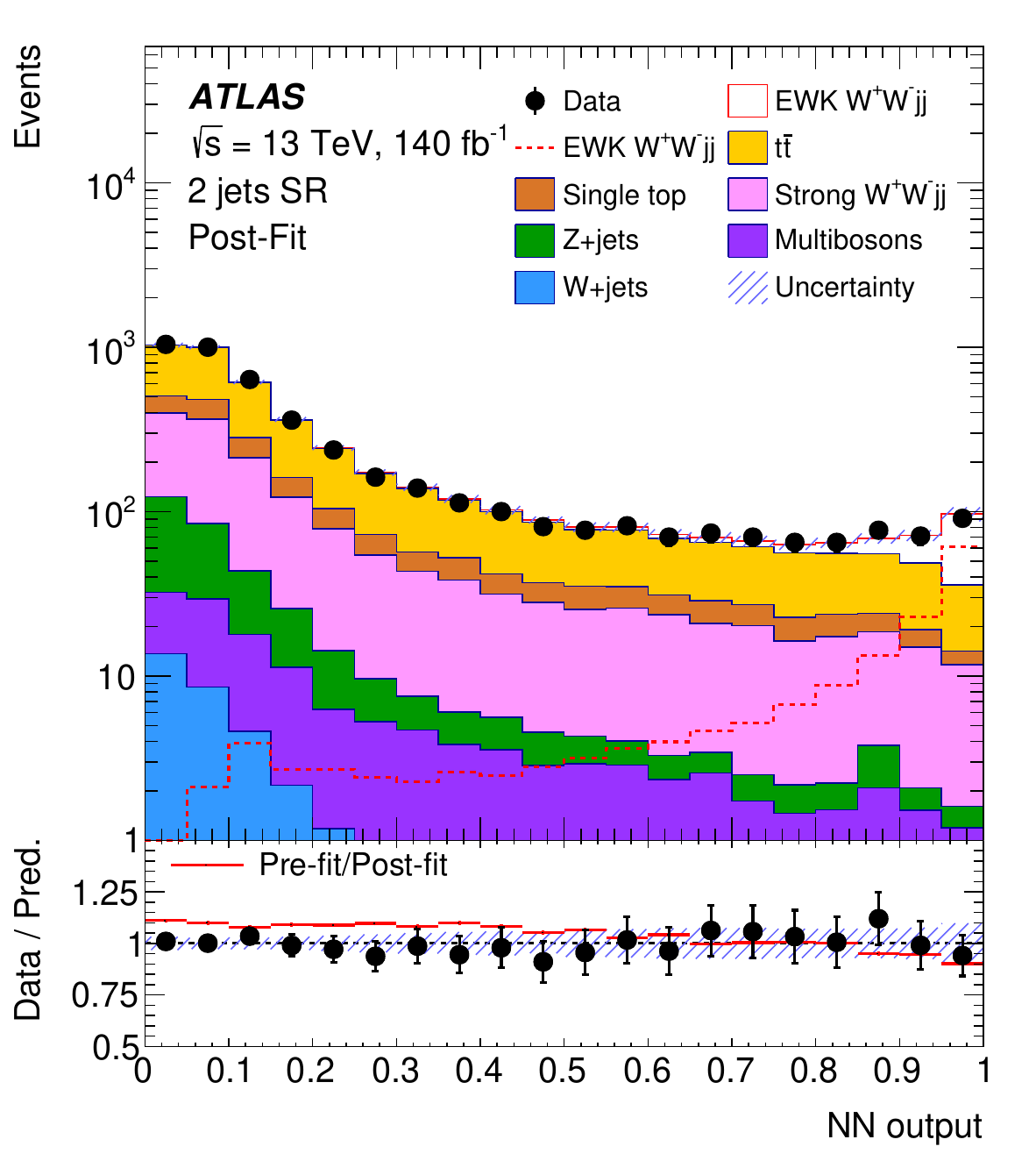}
}
\subfloat[]{
\includegraphics[width=0.45\textwidth]{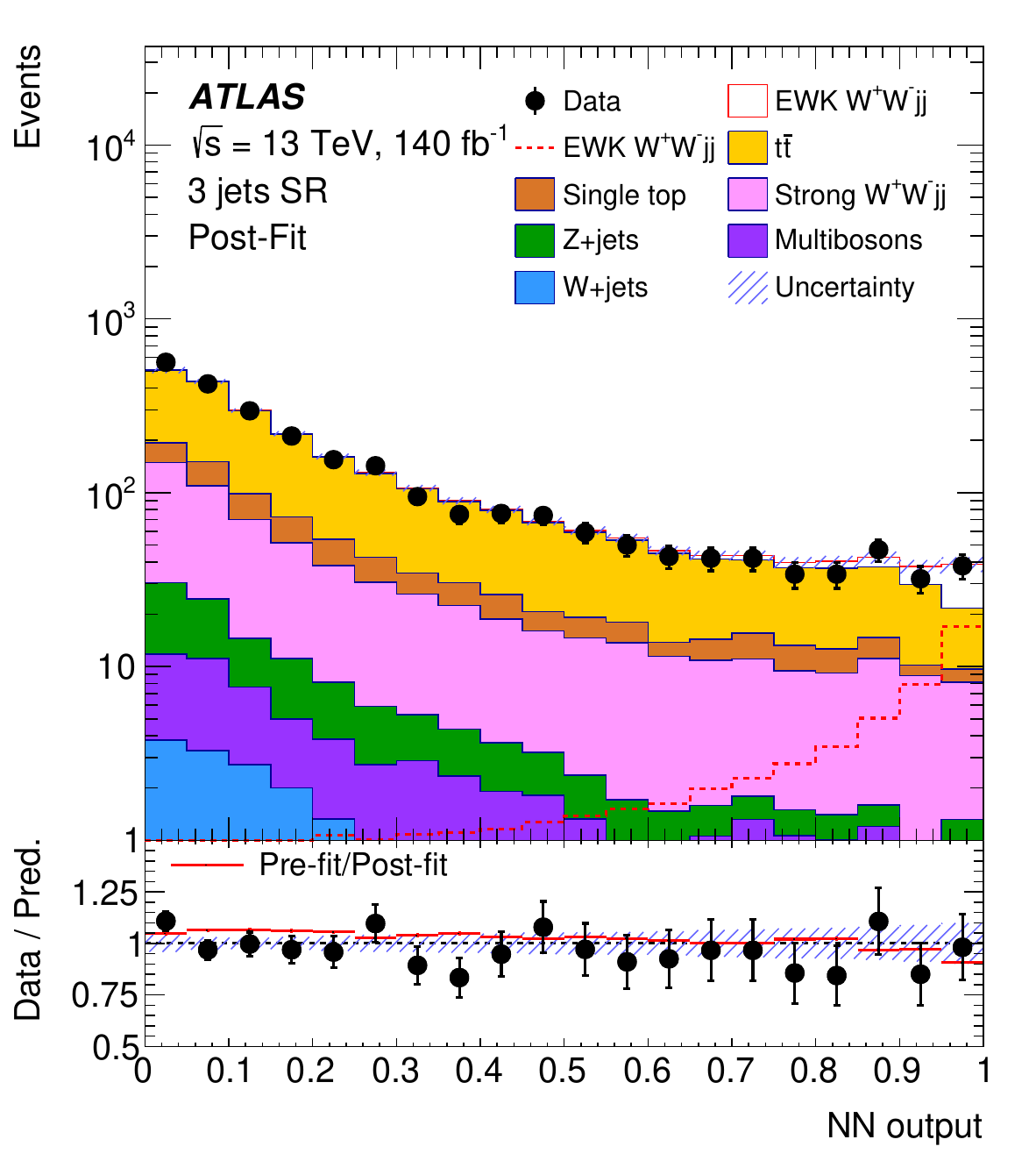}
}
\caption{ Distributions of the neural network output in the SR
for (a) two jets and (b) three jets, with the top quark, strong \WWosjj, and signal
processes constrained by the likelihood fit described in
Section~\ref{sec:result}.
Data corresponds to the filled dots, the signal normalised to its
simulated cross-section is represented by the unstacked dashed
histogram, and the post-fit predictions of the SM
together with the signal are depicted by the stacked filled
histograms.
The lower panels show the ratios of the data to the post-fit
predictions of the SM, and the ratio of the
pre-fit to the post-fit SM yields.
The uncertainty in the total contribution from the SM is
illustrated by a hashed band.
}
\label{fig:postfit_nn_sr}
\end{figure}

% The next lines are included from the .//sections/results.tex input file
A profile likelihood fit~\cite{Cranmer:1456844,JAMES1975343} is performed
on the NN output observable, simultaneously in the SR and CR,
with the normalisations of the signal ($\mu$), top quark and strong \WWosjj
processes set as floating parameters. These normalisations consist of
multiplicative factors to the nominal process simulation to match the
observation in data.
The NN is trained separately for events with two or three jets in the SR,
with distinct distributions being fitted for the two categories. However,
events with both jet multiplicities from the CR are merged together.
Figures~\ref{fig:postfit_nn_cr} and~\ref{fig:postfit_nn_sr} show the neural
network distribution after the fit in the CR and SR,
respectively. A good description of the data by the predicted signal and
background events can be observed in both figures.
 
\begin{figure}[!t]
\raggedleft
\includegraphics[width=0.75\textwidth]{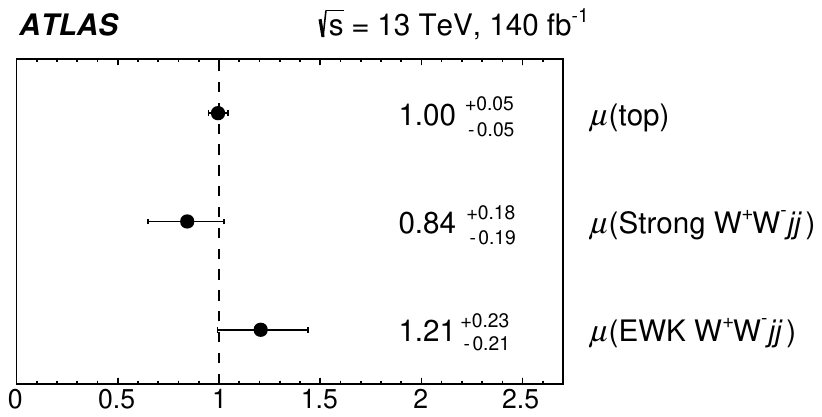}
\caption{ Measurement of the signal strength
$\mu=\mu(\mathrm{EWK}\;\WWosjj)$
from the likelihood fit described in Section~\ref{sec:result}
together with the normalisations of
the backgrounds originating from the top quark and strong
\WWosjj events. }
\label{fit:fit_norm_param}
\end{figure}

As illustrated in Figure~\ref{fit:fit_norm_param}, the normalisation of the
top quark background is well constrained by the fit, contrary to the
normalisation of the strong \WWosjj production, which does not benefit from
a dedicated CR, as it is challenging to isolate.
The impact of the uncertainty due to the normalisation of the
strong \WWosjj background on the total uncertainty is however limited, as
this process represents only about 10 to 15\% of the events at large NN
output values.
The split of the SR into a component with two jets and another one with
three jets was found to improve the expected signal significance by about
one standard deviation. 
The observed and expected
signal significance obtained from the fit are 7.1 and 6.2 standard deviations,
respectively.
 
The uncertainty in the signal strength $\mu$ derived from the likelihood fit
is referred to as $\Delta\mu$. For each source of systematic uncertainty,
a new fit is performed with the other nuisance parameters left constant at
their best fit values from the nominal fit, providing a variation of the uncertainty
in the signal strength $\Delta\mu^{'}$.
The impact of each category of uncertainty is defined as
$\sqrt{(\Delta\mu)^{2}-(\Delta\mu^{'})^{2}}/\mu$ and is reported in
Table~\ref{tab:sys_ranking_table}.
The uncertainty in the luminosity and the experimental uncertainties
specific to a given physics object are treated as fully correlated across the
different predictions and regions in the fit.
The theoretical uncertainties in the prediction of the top quark background are
considered as correlated between the CR and the SR.
The dominant uncertainty in the fit is due to the limited number of events in the data
and amounts to $12.3\%$, while the total uncertainty is $18.5\%$.
The nuisance parameters with an impact smaller than 0.5\% are pruned, with
negligible effect on the results.
 
\begin{table}[!t]
\centering
\caption{ Impact of systematic uncertainties on the signal strength $\mu$
after the fit. The different nuisance parameters are merged into
various categories. The statistical and total uncertainties are also
provided. }
\begin{tabular}{|l|S|}
\hline
Sources & \multicolumn{1}{c|}{$\frac{\sqrt{(\Delta\mu)^2 - (\Delta\mu')^2}}{\mu}$ [\%]} \\
\hline
\hline
MC statistical uncertainty &  7.7 \\
Top quark theoretical uncertainties &  6.3 \\
Signal theoretical uncertainties    &  5.8 \\
Jet experimental uncertainties      &  4.9 \\
Strong \WWosjj theoretical uncertainties &  1.3 \\
Luminosity                          &  0.8 \\
Misidentified lepton uncertainty   &  0.5 \\
$b$-tagging                         &  0.4 \\
Lepton experimental uncertainties   &  0.1 \\
Others                              &  0.3 \\
\hline
\hline
Data statistical uncertainty        & 12.3 \\ 
Top quark normalisation uncertainty &  4.9 \\
Strong \WWosjj normalisation uncertainty  &  2.2 \\
\hline
\hline
Total uncertainty                   & 18.5 \\ 
\hline
\end{tabular}
\medskip
\label{tab:sys_ranking_table}
\end{table}
 
The cross-section for VBS \WWosjj production is measured at particle level
in a fiducial region designed to be close to the most sensitive subset of
the SR\@.
Its event selection is detailed in Table~\ref{tab:fiducial_region},
and combines the two- and three-jet categories.
In the fiducial volume, particle-level electrons and muons are required to
originate from the hard scatter.
The impact of photons emitted in a cone of radius $\Delta R = 0.1$ around each
lepton direction is considered by adding their momenta to the lepton momentum.
At particle level, jets are defined by clustering stable final-state particles
using the anti-$k_{t}$ algorithm 
with a distance
parameter of $R=0.4$, and the missing transverse momentum is evaluated as the
transverse component of the vectorial sum of the neutrino momenta.
Events where the two leading jets have an
invariant mass $m_{jj}$ smaller than $500\,\GeV$ are excluded from the
fiducial region, while this is not required in the
reconstruction-level SR\@.
Events with $m_{jj}$ above $500\,\GeV$ in the fiducial region are primarily
reconstructed with NN output values larger than 0.6, which constitutes a
region where the fraction of reconstructed signal events in the SR exceeds
5\% in each NN bin and therefore drives the determination of the signal
strength $\mu$.
The cut on $m_{jj}$ in the fiducial volume allows the production
of triboson events 
to be suppressed, thus providing a fiducial cross-section purely related
to the \WWosjj production.
The fiducial cross-section for VBS \WWosjj production is obtained
by multiplying $\mu$ by the theoretical fiducial cross-section.
Therefore, the cut on $m_{jj}$ also avoids $\mu$ being applied to regions of
the phase space where the signal is negligible.

Using \POWHEGBOX[v2], 
the theoretical cross-section for the VBS $W^+W^-$
production is $2.20^{+0.14}_{-0.13}\,\mathrm{fb}$,
while the observed fiducial cross-sections is $2.65^{+0.49}_{-0.46}\,\mathrm{fb}$.

\begin{table}[!t]
\centering
\caption{Definition of the fiducial region at particle level.}
\begin{tabular}{ll}
\toprule
Category & Requirements \\
\midrule
Leptons  & $\pt>27\,\GeV$ and $|\eta|<2.5$ \smallskip \\
 
$b$-jets & $\pt>20\, \GeV$ and $|\eta| < 2.5$ \smallskip \\
Jets     & $\pt>25\, \GeV$ and $|\eta| < 4.5$ \\
\midrule
Events   & One electron and one muon with opposite electric charges \\
& No additional lepton \\
& $m_{e\mu} > 80\,\GeV$ \\
& $\met>15\,\GeV$ \\
& No $b$-jet \\
& Two or three jets \\
& $\zeta>0.5$ \\
& $m_{jj}>500\,\GeV$ \\
\bottomrule
\end{tabular}
\medskip
\label{tab:fiducial_region}
\end{table}

% End of text imported from the .//sections/results.tex input file

\FloatBarrier
 
\section{Conclusion}

% The next lines are included from the .//sections/conclusion.tex input file
The cross-section of the electroweak production of pairs of scattering $W$
bosons with opposite electric charges is measured using a dataset
amounting to $140\,\mathrm{fb}^{-1}$ of proton--proton collisions
at $\sqrt{s}=13\,\TeV$.
 
The analysis focuses on the final state consisting of an electron and a muon
with opposite electric charges, missing transverse energy, and two or three
jets. A NN is used to discriminate between the signal and the
dominant processes from the SM, namely top quark and strong
$W^{+}W^{-}jj$ production, and to extract the signal yield.
The signal is observed with a significance of $7.1$ standard deviations,
while $6.2$ standard deviations were expected.
The observed cross-section is determined in a
signal-enriched fiducial volume, and amounts to
$2.7\pm0.5\,\mathrm{fb}$, which is consistent with the theoretical prediction
of $2.20^{+0.14}_{-0.13}\text{ fb}$.
The dominant source of uncertainty is due to the limited number of events in the data.

% End of text imported from the .//sections/conclusion.tex input file

\section*{Acknowledgements}

% The next lines are included from the .//acknowledgements/Acknowledgements.tex input file

We thank CERN for the very successful operation of the LHC and its injectors, as well as the support staff at
CERN and at our institutions worldwide without whom ATLAS could not be operated efficiently.
 
The crucial computing support from all WLCG partners is acknowledged gratefully, in particular from CERN, the ATLAS Tier-1 facilities at TRIUMF/SFU (Canada), NDGF (Denmark, Norway, Sweden), CC-IN2P3 (France), KIT/GridKA (Germany), INFN-CNAF (Italy), NL-T1 (Netherlands), PIC (Spain), RAL (UK) and BNL (USA), the Tier-2 facilities worldwide and large non-WLCG resource providers. Major contributors of computing resources are listed in Ref.~\cite{ATL-SOFT-PUB-2023-001}.
 
We gratefully acknowledge the support of ANPCyT, Argentina; YerPhI, Armenia; ARC, Australia; BMWFW and FWF, Austria; ANAS, Azerbaijan; CNPq and FAPESP, Brazil; NSERC, NRC and CFI, Canada; CERN; ANID, Chile; CAS, MOST and NSFC, China; Minciencias, Colombia; MEYS CR, Czech Republic; DNRF and DNSRC, Denmark; IN2P3-CNRS and CEA-DRF/IRFU, France; SRNSFG, Georgia; BMBF, HGF and MPG, Germany; GSRI, Greece; RGC and Hong Kong SAR, China; ISF and Benoziyo Center, Israel; INFN, Italy; MEXT and JSPS, Japan; CNRST, Morocco; NWO, Netherlands; RCN, Norway; MEiN, Poland; FCT, Portugal; MNE/IFA, Romania; MESTD, Serbia; MSSR, Slovakia; ARIS and MVZI, Slovenia; DSI/NRF, South Africa; MICINN, Spain; SRC and Wallenberg Foundation, Sweden; SERI, SNSF and Cantons of Bern and Geneva, Switzerland; NSTC, Taipei; TENMAK, T\"urkiye; STFC/UKRI, United Kingdom; DOE and NSF, United States of America.
 
Individual groups and members have received support from BCKDF, CANARIE, CRC and DRAC, Canada; CERN-CZ, PRIMUS 21/SCI/017 and UNCE SCI/013, Czech Republic; COST, ERC, ERDF, Horizon 2020, ICSC-NextGenerationEU and Marie Sk{\l}odowska-Curie Actions, European Union; Investissements d'Avenir Labex, Investissements d'Avenir Idex and ANR, France; DFG and AvH Foundation, Germany; Herakleitos, Thales and Aristeia programmes co-financed by EU-ESF and the Greek NSRF, Greece; BSF-NSF and MINERVA, Israel; Norwegian Financial Mechanism 2014-2021, Norway; NCN and NAWA, Poland; La Caixa Banking Foundation, CERCA Programme Generalitat de Catalunya and PROMETEO and GenT Programmes Generalitat Valenciana, Spain; G\"{o}ran Gustafssons Stiftelse, Sweden; The Royal Society and Leverhulme Trust, United Kingdom.
 
In addition, individual members wish to acknowledge support from Chile: Agencia Nacional de Investigaci\'on y Desarrollo (FONDECYT 1190886, FONDECYT 1210400, FONDECYT 1230812, FONDECYT 1230987); China: National Natural Science Foundation of China (NSFC - 12175119, NSFC 12275265, NSFC-12075060); Czech Republic: PRIMUS Research Programme (PRIMUS/21/SCI/017); EU: H2020 European Research Council (ERC - 101002463); European Union: European Research Council (ERC - 948254), Horizon 2020 Framework Programme (MUCCA - CHIST-ERA-19-XAI-00), European Union, Future Artificial Intelligence Research (FAIR-NextGenerationEU PE00000013), Italian Center for High Performance Computing, Big Data and Quantum Computing (ICSC, NextGenerationEU), Marie Sklodowska-Curie Actions (EU H2020 MSC IF GRANT NO 101033496); France: Agence Nationale de la Recherche (ANR-20-CE31-0013, ANR-21-CE31-0013, ANR-21-CE31-0022, ANR-22-EDIR-0002), Investissements d'Avenir Idex (ANR-11-LABX-0012), Investissements d'Avenir Labex (ANR-11-LABX-0012); Germany: Baden-Württemberg Stiftung (BW Stiftung-Postdoc Eliteprogramme), Deutsche Forschungsgemeinschaft (DFG - CR 312/5-1); Italy: Istituto Nazionale di Fisica Nucleare (FELLINI G.A. n. 754496, ICSC, NextGenerationEU); Japan: Japan Society for the Promotion of Science (JSPS KAKENHI JP21H05085, JSPS KAKENHI JP22H01227, JSPS KAKENHI JP22H04944); Netherlands: Netherlands Organisation for Scientific Research (NWO Veni 2020 - VI.Veni.202.179); Norway: Research Council of Norway (RCN-314472); Poland: Polish National Agency for Academic Exchange (PPN/PPO/2020/1/00002/U/00001), Polish National Science Centre (NCN 2021/42/E/ST2/00350, NCN UMO-2019/34/E/ST2/00393, UMO-2020/37/B/ST2/01043, UMO-2021/40/C/ST2/00187); Slovenia: Slovenian Research Agency (ARIS grant J1-3010); Spain: BBVA Foundation (LEO22-1-603), Generalitat Valenciana (Artemisa, FEDER, IDIFEDER/2018/048), La Caixa Banking Foundation (LCF/BQ/PI20/11760025), Ministry of Science and Innovation (MCIN \& NextGenEU PCI2022-135018-2, MICIN \& FEDER PID2021-125273NB, RYC2019-028510-I, RYC2020-030254-I, RYC2021-031273-I, RYC2022-038164-I), PROMETEO and GenT Programmes Generalitat Valenciana (CIDEGENT/2019/023, CIDEGENT/2019/027); Sweden: Swedish Research Council (VR 2018-00482, VR 2022-03845, VR 2022-04683, VR grant 2021-03651), Knut and Alice Wallenberg Foundation (KAW 2017.0100, KAW 2018.0157, KAW 2018.0458, KAW 2019.0447); Switzerland: Swiss National Science Foundation (SNSF - PCEFP2\_194658); United Kingdom: Leverhulme Trust (Leverhulme Trust RPG-2020-004); United States of America: U.S. Department of Energy (ECA DE-AC02-76SF00515), Neubauer Family Foundation.

% End of text imported from the .//acknowledgements/Acknowledgements.tex input file

\printbibliography

@Article{alwall_automated_2014,
	title = {The automated computation of tree-level and next-to-leading order differential cross sections, and their matching to parton shower simulations},
	volume = {07},
	%issn = {1029-8479},
	%url = {http://arxiv.org/abs/1405.0301},
	doi = {10.1007/JHEP07(2014)079},
	number = {7},
	%urldate = {2022-10-05},
	journal = {JHEP},
	author = {Alwall, J. and Frederix, R. and Frixione, S. and Hirschi, V. and Maltoni, F. and Mattelaer, O. and Shao, H.-S. and Stelzer, T. and Torrielli, P. and Zaro, M.},
	month = jul,
        eprint = {1405.0301},
	primaryClass   = {hep-ph},
	archivePrefix  =  {arXiv},
        year = {2014},
	%note = {arXiv:1405.0301 [hep-ph]},
	keywords = {High Energy Physics - Phenomenology},
	pages = {79},
}

@article{Anastasiou_2004,
	doi = {10.1103/physrevd.69.094008},
%	url = {https://doi.org/10.1103%2Fphysrevd.69.094008}, 
	year = 2004,
%	month = {may}, 
%	publisher = {American Physical Society ({APS})},
	volume = {69},
	number = {9}, 
	author = {Charalampos Anastasiou and Lance Dixon and Kirill Melnikov and Frank Petriello},  
	title = {High-precision {QCD} at hadron colliders: Electroweak gauge boson rapidity distributions at next-to-next-to leading order},  
	journal = {Phys. Rev. D},
        eprint         = "hep-ph/0312266",
        archivePrefix  = "arXiv",
        %primaryClass   = "hep-ph",
}

@article{Caola_2016,
	doi = {10.1016/j.physletb.2016.01.046},
%	url = {https://doi.org/10.1016%2Fj.physletb.2016.01.046}, 
	year = 2016,
%	month = {mar}, 
%	publisher = {Elsevier {BV}}, 
	volume = {754},
	pages = {275--280},
	author = {Fabrizio Caola and Kirill Melnikov and Raoul Röntsch and Lorenzo Tancredi},
	title = {{QCD} corrections to \(W^+W^-\) production through gluon fusion},
	journal = {Phys. Lett. B}
}

@Article{MUON-2022-01,
%    author         = "{ATLAS Collaboration}",
%    title          = "{Studies of the muon momentum calibration and performance of the ATLAS detector with \(pp\) collisions at \(\sqrt{s} = 13\,\text{TeV}\)}",
%    year           = "2022",
%    reportNumber   = "CERN-EP-2022-274",
%    eprint         = "2212.07338",
%    archivePrefix  = "arXiv",
%    primaryClass   = "hep-ex",
%}

@article{Anastasiou:2015vya,
    author = "Anastasiou, Charalampos and Duhr, Claude and Dulat, Falko and Herzog, Franz and Mistlberger, Bernhard",
    title = "{Higgs Boson Gluon-Fusion Production in QCD at Three Loops}",
    eprint = "1503.06056",
    archivePrefix = "arXiv",
    primaryClass = "hep-ph",
    reportNumber = "CERN-PH-TH-2015-055, CP3-15-07",
    doi = "10.1103/PhysRevLett.114.212001",
    journal = "Phys. Rev. Lett.",
    volume = "114",
    pages = "212001",
    year = "2015"
}

@article{ballestrero_precise_2018,
	title = {Precise predictions for same-sign {W}-boson scattering at the {LHC}},
	volume = {78},
	%issn = {1434-6044, 1434-6052},
	%url = {http://arxiv.org/abs/1803.07943},
	doi = {10.1140/epjc/s10052-018-6136-y},
	abstract = {Vector-boson scattering processes are of great importance for the current run-II and future runs of the Large Hadron Collider. The presence of triple and quartic gauge couplings in the process gives access to the gauge sector of the Standard Model (SM) and possible new-physics contributions there. To test any new-physics hypothesis, sound knowledge of the SM contributions is necessary, with a precision which at least matches the experimental uncertainties of existing and forthcoming measurements. In this article we present a detailed study of the vector-boson scattering process with two positively-charged leptons and missing transverse momentum in the final state. In particular, we first carry out a systematic comparison of the various approximations that are usually performed for this kind of process against the complete calculation, at LO and NLO QCD accuracy. Such a study is performed both in the usual fiducial region used by experimental collaborations and in a more inclusive phase space, where the differences among the various approximations lead to more sizeable effects. Afterwards, we turn to predictions matched to parton showers, at LO and NLO: we show that on the one hand, the inclusion of NLO QCD corrections leads to more stable predictions, but on the other hand the details of the matching and of the parton-shower programs cause differences which are considerably larger than those observed at fixed order, even in the experimental fiducial region. We conclude with recommendations for experimental studies of vector-boson scattering processes.},
	number = {8},
	%urldate = {2023-09-29},
	journal = {Eur. Phys. J. C},
	author = {Ballestrero, Alessandro and Biedermann, Benedikt and Brass, Simon and Denner, Ansgar and Dittmaier, Stefan and Frederix, Rikkert and Govoni, Pietro and Grossi, Michele and Jäger, Barbara and Karlberg, Alexander and Maina, Ezio and Pellen, Mathieu and Pelliccioli, Giovanni and Plätzer, Simon and Rauch, Michael and Rebuzzi, Daniela and Reuter, Jürgen and Rothe, Vincent and Schwan, Christopher and Shao, Hua-Sheng and Stienemeier, Pascal and Zanderighi, Giulia and Zaro, Marco and Zeppenfeld, Dieter},
	month = aug,
	primaryClass   = {hep-ph},
	archivePrefix  =  {arXiv},
	year = {2018},
        eprint = {1803.07943},
        %note = {arXiv:1803.07943 [hep-ph]},
	keywords = {High Energy Physics - Phenomenology},
	pages = {671},
	annote = {Comment: Data and input files are available at: https://github.com/VBSCanAction/VBSComparison/ . 28 pages including figures and references. Matches the journal version},
}

@Article{Cacciari:2008gp,
     author    = "Cacciari, Matteo and Salam, Gavin P. and Soyez, Gregory",
     title     = "{The anti-\(k_{t}\) jet clustering algorithm}",
     journal   = "JHEP",
     volume    = "04",
     year      = "2008",
     pages     = "063",
     eprint    = "0802.1189",
     archivePrefix = "arXiv",
     primaryClass  =  "hep-ph",
     doi       = "10.1088/1126-6708/2008/04/063",
     SLACcitation  = "%%CITATION = 0802.1189;%%"
}

@Article{Butterworth:2015oua,
      author         = "Butterworth, Jon and others",
      title          = "{PDF4LHC recommendations for LHC Run II}",
      journal        = "J. Phys. G",
      volume         = "43",
      year           = "2016",
      pages          = "023001",
      doi            = "10.1088/0954-3899/43/2/023001",
      eprint         = "1510.03865",
      archivePrefix  = "arXiv",
      primaryClass   = "hep-ph",
      reportNumber   = "OUTP-15-17P, SMU-HEP-15-12, TIF-UNIMI-2015-14,
                        LCTS-2015-27, CERN-PH-TH-2015-249",
      SLACcitation   = "%%CITATION = ARXIV:1510.03865;%%"
}

@Article{Pumplin:2002vw,
     author    = "Pumplin, J. and others",
     title     = "{New Generation of Parton Distributions with Uncertainties from Global QCD Analysis}",
     journal   = "JHEP",
     volume    = "07",
     year      = "2002",
     pages     = "012",
     doi       = "10.1088/1126-6708/2002/07/012",
     eprint    = "hep-ph/0201195",
     archivePrefix = "arXiv",
     SLACcitation  = "%%CITATION = HEP-PH/0201195;%%"
}

@article{Harland-Lang:2014zoa,
      author         = "Harland-Lang, L. A. and Martin, A. D. and Motylinski, P.
                        and Thorne, R. S.",
      title          = "{Parton distributions in the LHC era: MMHT 2014 PDFs}",
      journal        = "Eur. Phys. J. C",
      number         = "5",
      volume         = "75",
      pages          = "204",
      doi            = "10.1140/epjc/s10052-015-3397-6",
      year           = "2015",
      eprint         = "1412.3989",
      archivePrefix  = "arXiv",
      primaryClass   = "hep-ph",
      reportNumber   = "LCTS-2014-47, IPPP-14-97, DCPT-14-194",
      SLACcitation   = "%%CITATION = ARXIV:1412.3989;%%",
}

@article{Ball:2012cx,
      author         = "{NNPDF Collaboration} and Ball, Richard D. and others",
      title          = "{Parton distributions with LHC data}",
      journal        = "Nucl. Phys. B",
      volume         = "867",
      year           = "2013",
      pages          = "244",
      doi            = "10.1016/j.nuclphysb.2012.10.003",
      eprint         = "1207.1303",
      archivePrefix  = "arXiv",
      primaryClass   = "hep-ph",
      reportNumber   = "EDINBURGH-2012-08, IFUM-FT-997, FR-PHENO-2012-014,
                        RWTH-TTK-12-25, CERN-PH-TH-2012-037, SFB-CPP-12-47\,
                        --CERN-PH-TH-2012-037",
      SLACcitation   = "%%CITATION = ARXIV:1207.1303;%%"
}

@article{Ball:2014uwa,
      author         = "{NNPDF Collaboration} and Ball, Richard D. and others",
      title          = "{Parton distributions for the LHC run II}",
      journal        = "JHEP",
      volume         = "04",
      year           = "2015",
      pages          = "040",
      doi            = "10.1007/JHEP04(2015)040",
      eprint         = "1410.8849",
      archivePrefix  = "arXiv",
      primaryClass   = "hep-ph",
      reportNumber   = "EDINBURGH-2014-15, IFUM-1034-FT, CERN-PH-TH-2013-253,
                        OUTP-14-11P, CAVENDISH-HEP-14-11",
      SLACcitation   = "%%CITATION = ARXIV:1410.8849;%%"
}

@article{Czakon:2012pz,
      author         = "Czakon, Michal and Mitov, Alexander",
      title          = "{NNLO corrections to top pair production at hadron
                        colliders: the quark-gluon reaction}",
      journal        = "JHEP",
      volume         = "01",
      pages          = "080",
      doi            = "10.1007/JHEP01(2013)080",
      year           = "2013",
      eprint         = "1210.6832",
      archivePrefix  = "arXiv",
      primaryClass   = "hep-ph",
      SLACcitation   = "%%CITATION = ARXIV:1210.6832;%%",
}

@Article{Sjostrand:2007gs,
	author    = "Sj{\"o}strand, T. and Mrenna, S. and Skands, P.",
	title     = "{A brief introduction to PYTHIA 8.1}",
	journal   = "Comput. Phys. Commun.",
	volume    = "178",
	year      = "2008",
	pages     = "852-867",
	eprint    = "0710.3820",
	archivePrefix = "arXiv",
	primaryClass  =  "hep-ph",
	doi       = "10.1016/j.cpc.2008.01.036",
	SLACcitation  = "%%CITATION = 0710.3820;%%"
}

@article{Sjostrand:2014zea,
      author         = "Sj{\"o}strand, Torbj{\"o}rn and Ask, Stefan and Christiansen,
                        Jesper R. and Corke, Richard and Desai, Nishita and Ilten,
                        Philip and Mrenna, Stephen and Prestel, Stefan and
                        Rasmussen, Christine O. and Skands, Peter Z.",
      title          = "{An introduction to PYTHIA 8.2}",
      journal        = "Comput. Phys. Commun.",
      volume         = "191",
      year           = "2015",
      pages          = "159",
      doi            = "10.1016/j.cpc.2015.01.024",
      eprint         = "1410.3012",
      archivePrefix  = "arXiv",
      primaryClass   = "hep-ph",
      reportNumber   = "LU-TP-14-36, MCNET-14-22, CERN-PH-TH-2014-190,
                        FERMILAB-PUB-14-316-CD, DESY-14-178, SLAC-PUB-16122,
                        --FERMILAB-PUB-14-316-CD",
      SLACcitation   = "%%CITATION = ARXIV:1410.3012;%%"
}

@Article{Lange:2001uf,
      author         = "Lange, D. J.",
      title          = "{The EvtGen particle decay simulation package}",
      booktitle      = "{Proceedings, 7th International Conference on B physics
                        at hadron machines (BEAUTY 2000)}",
      journal        = "Nucl. Instrum. Meth. A",
      volume         = "462",
      year           = "2001",
      pages          = "152",
      doi            = "10.1016/S0168-9002(01)00089-4",
      SLACcitation   = "%%CITATION = NUIMA,A462,152;%%"
}

@Article{Frixione:2008yi,
     author    = "Frixione, Stefano and Laenen, Eric and Motylinski, Patrick and White, Chris and Webber, Bryan R.",
     title     = "{Single-top hadroproduction in association with a \(W\) boson}",
     journal   = "JHEP",
     volume    = "07",
     year      = "2008",
     pages     = "029",
     eprint    = "0805.3067",
     archivePrefix = "arXiv",
     primaryClass  =  "hep-ph",
     doi       = "10.1088/1126-6708/2008/07/029"
}

@Article{Bahr:2008pv,
      author         = "B{\"a}hr, M. and others",
      title          = "{Herwig++ physics and manual}",
      journal        = "Eur. Phys. J. C",
      volume         = "58",
      year           = "2008",
      pages          = "639",
      doi            = "10.1140/epjc/s10052-008-0798-9",
      eprint         = "0803.0883",
      archivePrefix  = "arXiv",
      primaryClass   = "hep-ph",
      reportNumber   = "CERN-PH-TH-2008-038, CAVENDISH-HEP-08-03, KA-TP-05-2008,
                        DCPT-08-22, IPPP-08-11, CP3-08-05",
      SLACcitation   = "%%CITATION = ARXIV:0803.0883;%%"
}

@Article{Bellm:2015jjp,
      author         = "Bellm, Johannes and others",
      title          = "{Herwig 7.0/Herwig++ 3.0 release note}",
      journal        = "Eur. Phys. J. C",
      volume         = "76",
      year           = "2016",
      number         = "4",
      pages          = "196",
      doi            = "10.1140/epjc/s10052-016-4018-8",
      eprint         = "1512.01178",
      archivePrefix  = "arXiv",
      primaryClass   = "hep-ph",
      reportNumber   = "CERN-PH-TH-2015-289, MAN-HEP-2015-15, IFJPAN-IV-2015-13,
                        HERWIG-2015-01, KA-TP-18-2015, DCPT-15-142, MCNET-15-28,
                        IPPP-15-71, --HERWIG-2015-01",
      SLACcitation   = "%%CITATION = ARXIV:1512.01178;%%"
}

@Article{Nason:2004rx,
      author         = "Nason, Paolo",
      title          = "{A new method for combining NLO QCD with shower Monte Carlo algorithms}",
      journal        = "JHEP",
      volume         = "11",
      pages          = "040",
      doi            = "10.1088/1126-6708/2004/11/040",
      year           = "2004",
      eprint         = "hep-ph/0409146",
      archivePrefix  = "arXiv",
}

@Article{Frixione:2007vw,
      author         = "Frixione, Stefano and Nason, Paolo and Oleari, Carlo",
      title          = "{Matching NLO QCD computations with parton shower
                        simulations: the POWHEG method}",
      journal        = "JHEP",
      volume         = "11",
      pages          = "070",
      doi            = "10.1088/1126-6708/2007/11/070",
      year           = "2007",
      eprint         = "0709.2092",
      archivePrefix  = "arXiv",
      primaryClass   = "hep-ph",
}

@Article{Alioli:2010xd,
      author         = "Alioli, Simone and Nason, Paolo and Oleari, Carlo and Re,
                        Emanuele",
      title          = "{A general framework for implementing NLO calculations in
                        shower Monte Carlo programs: the POWHEG BOX}",
      journal        = "JHEP",
      volume         = "06",
      pages          = "043",
      doi            = "10.1007/JHEP06(2010)043",
      year           = "2010",
      eprint         = "1002.2581",
      archivePrefix  = "arXiv",
      primaryClass   = "hep-ph",
}

@article{Kidonakis:2010ux,
      author         = "Kidonakis, Nikolaos",
      title          = "{Two-loop soft anomalous dimensions for single top quark
                        associated production with a \(W^{-}\) or \(H^{-}\)}",
      journal        = "Phys. Rev. D",
      volume         = "82",
      year           = "2010",
      pages          = "054018",
      doi            = "10.1103/PhysRevD.82.054018",
      eprint         = "1005.4451",
      archivePrefix  = "arXiv",
      primaryClass   = "hep-ph",
      SLACcitation   = "%%CITATION = ARXIV:1005.4451;%%"
}

@inproceedings{Kidonakis:2013zqa,
      author         = "Kidonakis, Nikolaos",
      title          = "{Top Quark Production}",
      booktitle      = "{Proceedings, Helmholtz International Summer School on
                        Physics of Heavy Quarks and Hadrons (HQ 2013)}",
      eventdate      = "2013-07-15/2013-07-28",
      venue          = "JINR, Dubna, Russia",
      pages          = "139-168",
      doi            = "10.3204/DESY-PROC-2013-03/Kidonakis",
      eprint         = "1311.0283",
      archivePrefix  = "arXiv",
      primaryClass   = "hep-ph",
      reportNumber   = "KSU-HEP-110113",
      SLACcitation   = "%%CITATION = ARXIV:1311.0283;%%"
}

@article{Beneke:2011mq,
      author         = "Beneke, M. and Falgari, P. and Klein, S. and Schwinn, C.",
      title          = "{Hadronic top-quark pair production with NNLL threshold
                        resummation}",
      journal        = "Nucl. Phys. B",
      volume         = "855",
      year           = "2012",
      pages          = "695-741",
      doi            = "10.1016/j.nuclphysb.2011.10.021",
      eprint         = "1109.1536",
      archivePrefix  = "arXiv",
      primaryClass   = "hep-ph",
      reportNumber   = "TTK-11-38, ITP-UU-11-26, SPIN-11-19, FR-PHENO-2011-015,
                        SFB-CPP-11-49",
      SLACcitation   = "%%CITATION = ARXIV:1109.1536;%%"
}

@article{Cacciari:2011hy,
      author         = "Cacciari, Matteo and Czakon, Michal and Mangano,
                        Michelangelo and Mitov, Alexander and Nason, Paolo",
      title          = "{Top-pair production at hadron colliders with
                        next-to-next-to-leading logarithmic soft-gluon
                        resummation}",
      journal        = "Phys. Lett. B",
      volume         = "710",
      pages          = "612-622",
      doi            = "10.1016/j.physletb.2012.03.013",
      year           = "2012",
      eprint         = "1111.5869",
      primaryClass   = "hep-ph",
      archivePrefix  = "arXiv",
      reportNumber   = "CERN-PH-TH-2011-277, TTK-11-54",
      SLACcitation   = "%%CITATION = ARXIV:1111.5869;%%",
}

@article{Czakon:2012zr,
      author         = "Czakon, Michal and Mitov, Alexander",
      title          = "{NNLO corrections to top-pair production at hadron
                        colliders: the all-fermionic scattering channels}",
      journal        = "JHEP",
      volume         = "12",
      pages          = "054",
      doi            = "10.1007/JHEP12(2012)054",
      year           = "2012",
      eprint         = "1207.0236",
      archivePrefix  = "arXiv",
      primaryClass   = "hep-ph",
      SLACcitation   = "%%CITATION = ARXIV:1207.0236;%%",
}

@article{Czakon:2013goa,
      author         = "Czakon, Michal and Fiedler, Paul and Mitov, Alexander",
      title          = "{Total Top-Quark Pair-Production Cross Section at
                        Hadron Colliders Through \(O(\alpha_S^4)\)}",
      journal        = "Phys. Rev. Lett.",
      volume         = "110",
      pages          = "252004",
      doi            = "10.1103/PhysRevLett.110.252004",
      year           = "2013",
      eprint         = "1303.6254",
      archivePrefix  = "arXiv",
      primaryClass   = "hep-ph",
      reportNumber   = "CERN-PH-TH-2013-056, TTK-13-08",
      SLACcitation   = "%%CITATION = ARXIV:1303.6254;%%",
}

@article{Baernreuther:2012ws,
      author         = "B{\"a}rnreuther, Peter and Czakon, Michal and Mitov,
                        Alexander",
      title          = "{Percent-Level-Precision Physics at the
                  Tevatron: Next-to-Next-to-Leading Order QCD
                  Corrections to \(q \bar{q} \to t \bar{t} + X\)}",
      journal        = "Phys. Rev. Lett.",
      volume         = "109",
      pages          = "132001",
      doi            = "10.1103/PhysRevLett.109.132001",
      year           = "2012",
      eprint         = "1204.5201",
      archivePrefix  = "arXiv",
      primaryClass   = "hep-ph",
      SLACcitation   = "%%CITATION = ARXIV:1204.5201;%%",
}

@article{Frixione:2007nw,
      author         = "Frixione, Stefano and Ridolfi, Giovanni and Nason, Paolo",
      title          = "{A positive-weight next-to-leading-order Monte Carlo for
                        heavy flavour hadroproduction}",
      journal        = "JHEP",
      volume         = "09",
      pages          = "126",
      doi            = "10.1088/1126-6708/2007/09/126",
      year           = "2007",
      eprint         = "0707.3088",
      archivePrefix  = "arXiv",
      primaryClass   = "hep-ph",
      SLACcitation   = "%%CITATION = ARXIV:0707.3088;%%",
}

@article{deFlorian:2016spz,
      author         = "de Florian, D. and others",
      title          = "{Handbook of LHC Higgs Cross Sections: 4. Deciphering the
                        Nature of the Higgs Sector}",
      collaboration  = "LHC Higgs Cross Section Working Group",
      doi            = "10.23731/CYRM-2017-002",
      year           = "2016",
      eprint         = "1610.07922",
      archivePrefix  = "arXiv",
      primaryClass   = "hep-ph",
      reportNumber   = "FERMILAB-FN-1025-T, CERN-2017-002-M",
      SLACcitation   = "%%CITATION = ARXIV:1610.07922;%%"
}

@Article{Hoeche:2009rj,
      author         = "H{\"o}che, Stefan and Krauss, Frank and Schumann, Steffen and Siegert, Frank",
      title          = "{QCD matrix elements and truncated showers}",
      journal        = "JHEP",
      volume         = "05",
      pages          = "053",
      doi            = "10.1088/1126-6708/2009/05/053",
      year           = "2009",
      eprint         = "0903.1219",
      archivePrefix  = "arXiv",
      primaryClass   = "hep-ph",
}

@Article{Gleisberg:2008fv,
      author         = "Gleisberg, Tanju and H{\"o}che, Stefan",
      title          = "{Comix, a new matrix element generator}",
      journal        = "JHEP",
      volume         = "12",
      pages          = "039",
      doi            = "10.1088/1126-6708/2008/12/039",
      year           = "2008",
      eprint         = "0808.3674",
      archivePrefix  = "arXiv",
      primaryClass   = "hep-ph",
}

@Article{Schumann:2007mg,
      author         = "Schumann, Steffen and Krauss, Frank",
      title          = "{A parton shower algorithm based on Catani--Seymour dipole factorisation}",
      journal        = "JHEP",
      volume         = "03",
      pages          = "038",
      doi            = "10.1088/1126-6708/2008/03/038",
      year           = "2008",
      eprint         = "0709.1027",
      archivePrefix  = "arXiv",
      primaryClass   = "hep-ph",
}

@article{Bothmann:2019yzt,
      author = "Bothmann, Enrico and others",
      title = "{Event generation with Sherpa 2.2}",
      journal = "SciPost Phys.",
      volume = "7",
      year = "2019",
      number = "3",
      pages = "034",
      doi = "10.21468/SciPostPhys.7.3.034",
      reportNumber = "FERMILAB-PUB-19-218-T, SLAC-PUB-17433, IPPP/19/42, MCNET-19-11",
      eprint = "1905.09127",
      archivePrefix = "arXiv",
      primaryClass = "hep-ph",
}

@article{Hoeche:2011fd,
      author         = "H{\"o}che, Stefan and Krauss, Frank and Sch{\"o}nherr, Marek and
                        Siegert, Frank",
      title          = "{A critical appraisal of NLO+PS matching methods}",
      journal        = "JHEP",
      volume         = "09",
      year           = "2012",
      pages          = "049",
      doi            = "10.1007/JHEP09(2012)049",
      eprint         = "1111.1220",
      archivePrefix  = "arXiv",
      primaryClass   = "hep-ph",
      reportNumber   = "SLAC-PUB-14661, IPPP-11-67, DCPT-11-134, LPN11-58,
                        FR-PHENO-2011-019, MCNET-11-24",
      SLACcitation   = "%%CITATION = ARXIV:1111.1220;%%"
}

@article{Catani:2001cc,
      author         = "Catani, S. and Krauss, F. and Webber, B. R. and Kuhn, R.",
      title          = "{QCD Matrix Elements + Parton Showers}",
      journal        = "JHEP",
      volume         = "11",
      year           = "2001",
      pages          = "063",
      doi            = "10.1088/1126-6708/2001/11/063",
      eprint         = "hep-ph/0109231",
      archivePrefix  = "arXiv",
      reportNumber   = "CERN-TH-2000-367, CAVENDISH-HEP-00-03",
      SLACcitation   = "%%CITATION = HEP-PH/0109231;%%"
}

@article{Cascioli:2011va,
      author         = "Cascioli, Fabio and Maierh{\"o}fer, Philipp and Pozzorini, Stefano",
      title          = "{Scattering Amplitudes with Open Loops}",
      journal        = "Phys. Rev. Lett.",
      volume         = "108",
      year           = "2012",
      pages          = "111601",
      doi            = "10.1103/PhysRevLett.108.111601",
      eprint         = "1111.5206",
      archivePrefix  = "arXiv",
      primaryClass   = "hep-ph",
      reportNumber   = "ZU-TH-23-11, LPN11-66",
      SLACcitation   = "%%CITATION = ARXIV:1111.5206;%%"
}

@article{Denner:2016kdg,
      author         = "Denner, Ansgar and Dittmaier, Stefan and Hofer, Lars",
      title          = "{\textsc{Collier}: A fortran-based complex one-loop library in
                        extended regularizations}",
      journal        = "Comput. Phys. Commun.",
      volume         = "212",
      year           = "2017",
      pages          = "220-238",
      doi            = "10.1016/j.cpc.2016.10.013",
      eprint         = "1604.06792",
      archivePrefix  = "arXiv",
      primaryClass   = "hep-ph",
      reportNumber   = "FR-PHENO-2016-003, ICCUB-16-016",
      SLACcitation   = "%%CITATION = ARXIV:1604.06792;%%"
}

@article{Czakon:2011xx,
  author         = "Czakon, Michal and Mitov, Alexander",
  title          = "{Top++: A program for the calculation of the top-pair cross-section at hadron colliders}",
  journal        = "Comput. Phys. Commun.",
  volume         = "185",
  year           = "2014",
  pages          = "2930",
  doi            = "10.1016/j.cpc.2014.06.021",
  eprint         = "1112.5675",
  archivePrefix  = "arXiv",
  primaryClass   = "hep-ph",
  reportNumber   = "CERN-PH-TH-2011-303, TTK-11-58",
}

@article{Hamilton:2013fea,
      author         = "Hamilton, Keith and Nason, Paolo and Re, Emanuele and
                        Zanderighi, Giulia",
      title          = "{NNLOPS simulation of Higgs boson production}",
      journal        = "JHEP",
      volume         = "10",
      year           = "2013",
      pages          = "222",
      doi            = "10.1007/JHEP10(2013)222",
      eprint         = "1309.0017",
      archivePrefix  = "arXiv",
      primaryClass   = "hep-ph",
      reportNumber   = "MCNET-13-11, CERN-PH-TH-2013-205, OUTP-13-18P",
      SLACcitation   = "%%CITATION = ARXIV:1309.0017;%%"
}

@article{Hamilton:2015nsa,
      author         = "Hamilton, Keith and Nason, Paolo and Zanderighi, Giulia",
      title          = "{Finite quark-mass effects in the NNLOPS POWHEG+MiNLO
                        Higgs generator}",
      journal        = "JHEP",
      volume         = "05",
      year           = "2015",
      pages          = "140",
      doi            = "10.1007/JHEP05(2015)140",
      eprint         = "1501.04637",
      archivePrefix  = "arXiv",
      primaryClass   = "hep-ph",
      reportNumber   = "CERN-PH-TH-2015-006",
      SLACcitation   = "%%CITATION = ARXIV:1501.04637;%%"
}

@article{Anastasiou:2016cez,
      author         = "Anastasiou, Charalampos and Duhr, Claude and Dulat, Falko
                        and Furlan, Elisabetta and Gehrmann, Thomas and Herzog,
                        Franz and Lazopoulos, Achilleas and Mistlberger, Bernhard",
      title          = "{High precision determination of the gluon fusion Higgs
                        boson cross-section at the LHC}",
      journal        = "JHEP",
      volume         = "05",
      year           = "2016",
      pages          = "058",
      doi            = "10.1007/JHEP05(2016)058",
      eprint         = "1602.00695",
      archivePrefix  = "arXiv",
      primaryClass   = "hep-ph",
      reportNumber   = "CP3-16-01, ZU-TH-27-15, NIKHEF-2016-004,
                        CERN-TH-2016-006",
      SLACcitation   = "%%CITATION = ARXIV:1602.00695;%%"
}

@article{Dulat:2018rbf,
      author         = "Dulat, Falko and Lazopoulos, Achilleas and Mistlberger,
                        Bernhard",
      title          = "{iHixs 2 -- Inclusive Higgs cross sections}",
      journal        = "Comput. Phys. Commun.",
      volume         = "233",
      year           = "2018",
      pages          = "243-260",
      doi            = "10.1016/j.cpc.2018.06.025",
      eprint         = "1802.00827",
      archivePrefix  = "arXiv",
      primaryClass   = "hep-ph",
      reportNumber   = "CERN-TH-2018-019, SLAC-PUB-17222",
      SLACcitation   = "%%CITATION = ARXIV:1802.00827;%%"
}

@article{Harlander:2009mq,
      author         = "Harlander, Robert V. and Ozeren, Kemal J.",
      title          = "{Finite top mass effects for hadronic Higgs production at
                        next-to-next-to-leading order}",
      journal        = "JHEP",
      volume         = "11",
      year           = "2009",
      pages          = "088",
      doi            = "10.1088/1126-6708/2009/11/088",
      eprint         = "0909.3420",
      archivePrefix  = "arXiv",
      primaryClass   = "hep-ph",
      reportNumber   = "WUB-09-10",
      SLACcitation   = "%%CITATION = ARXIV:0909.3420;%%"
}

@article{Harlander:2009bw,
      author         = "Harlander, Robert V. and Ozeren, Kemal J.",
      title          = "{Top mass effects in Higgs production at
                        next-to-next-to-leading order QCD: Virtual corrections}",
      journal        = "Phys. Lett. B",
      volume         = "679",
      year           = "2009",
      pages          = "467-472",
      doi            = "10.1016/j.physletb.2009.08.012",
      eprint         = "0907.2997",
      archivePrefix  = "arXiv",
      primaryClass   = "hep-ph",
      reportNumber   = "WUB-09-08",
      SLACcitation   = "%%CITATION = ARXIV:0907.2997;%%"
}

@article{Harlander:2009my,
      author         = "Harlander, Robert V. and Mantler, Hendrik and Marzani,
                        Simone and Ozeren, Kemal J.",
      title          = "{Higgs production in gluon fusion at
                        next-to-next-to-leading order QCD for finite top mass}",
      journal        = "Eur. Phys. J. C",
      volume         = "66",
      year           = "2010",
      pages          = "359-372",
      doi            = "10.1140/epjc/s10052-010-1258-x",
      eprint         = "0912.2104",
      archivePrefix  = "arXiv",
      primaryClass   = "hep-ph",
      reportNumber   = "WUB-09-18, MAN-HEP-2009-44",
      SLACcitation   = "%%CITATION = ARXIV:0912.2104;%%"
}

@article{Pak:2009dg,
      author         = "Pak, Alexey and Rogal, Mikhail and Steinhauser, Matthias",
      title          = "{Finite top quark mass effects in NNLO Higgs boson
                        production at LHC}",
      journal        = "JHEP",
      volume         = "02",
      year           = "2010",
      pages          = "025",
      doi            = "10.1007/JHEP02(2010)025",
      eprint         = "0911.4662",
      archivePrefix  = "arXiv",
      primaryClass   = "hep-ph",
      reportNumber   = "SFB-CPP-09-116, TTP09-43",
      SLACcitation   = "%%CITATION = ARXIV:0911.4662;%%"
}

@article{Actis:2008ug,
      author         = "Actis, Stefano and Passarino, Giampiero and Sturm,
                        Christian and Uccirati, Sandro",
      title          = "{NLO electroweak corrections to Higgs boson production at
                        hadron colliders}",
      journal        = "Phys. Lett. B",
      volume         = "670",
      year           = "2008",
      pages          = "12-17",
      doi            = "10.1016/j.physletb.2008.10.018",
      eprint         = "0809.1301",
      archivePrefix  = "arXiv",
      primaryClass   = "hep-ph",
      reportNumber   = "PITHA-08-20, SFB-CPP-08-62, TTP08-38",
      SLACcitation   = "%%CITATION = ARXIV:0809.1301;%%"
}

@article{Actis:2008ts,
      author         = "Actis, Stefano and Passarino, Giampiero and Sturm,
                        Christian and Uccirati, Sandro",
      title          = "{NNLO computational techniques: The cases \(H \to \gamma\gamma\) and \(H \to gg\)}",
      journal        = "Nucl. Phys. B",
      volume         = "811",
      year           = "2009",
      pages          = "182-273",
      doi            = "10.1016/j.nuclphysb.2008.11.024",
      eprint         = "0809.3667",
      archivePrefix  = "arXiv",
      primaryClass   = "hep-ph",
      reportNumber   = "PITHA-08-24, SFB-CPP-08-73, TTP08-42",
      SLACcitation   = "%%CITATION = ARXIV:0809.3667;%%"
}

@article{Bonetti:2018ukf,
      author         = "Bonetti, Marco and Melnikov, Kirill and Tancredi,
                        Lorenzo",
      title          = "{Higher order corrections to mixed QCD-EW contributions
                        to Higgs boson production in gluon fusion}",
      journal        = "Phys. Rev. D",
      volume         = "97",
      year           = "2018",
      number         = "5",
      pages          = "056017",
      doi            = "10.1103/PhysRevD.97.056017",
      eprint         = "1801.10403",
      archivePrefix  = "arXiv",
      primaryClass   = "hep-ph",
      reportNumber   = "CERN-TH-2018-011, TTP18-004",
      SLACcitation   = "%%CITATION = ARXIV:1801.10403;%%",
      related = "Bonetti:2018ukf-err",
      relatedstring  = "Erratum:",
}

@article{Agostinelli:2002hh,
      author         = "Agostinelli, S. and others",
      title          = "{\textsc{Geant4} -- a simulation toolkit}",
      journal        = "Nucl. Instrum. Meth. A",
      volume         = "506",
      year           = "2003",
      pages          = "250",
      doi            = "10.1016/S0168-9002(03)01368-8",
      reportNumber   = "SLAC-PUB-9350, FERMILAB-PUB-03-339",
      SLACcitation   = "%%CITATION = NUIMA,A506,250;%%"
}

@Booklet{TMVA,
      author         = "Hoecker, Andreas and others",
      title          = "{TMVA - Toolkit for Multivariate Data Analysis}",
      year           = "2007",
      eprint         = "physics/0703039",
      archivePrefix  = "arXiv",
      primaryClass   = "physics.data-an",
}

@online{chollet2015keras,
      author         = {Chollet, Fran\c{c}ois and others},
      title          = {Keras},
      year           = {2015},
      url            = {https://keras.io},
}

@Article{PERF-2007-01,
    author         = "{ATLAS Collaboration}",
    title          = "{The ATLAS Experiment at the CERN Large Hadron Collider}",
    journal        = "JINST",
    volume         = "3",
    year           = "2008",
    pages          = "S08003",
    doi            = "10.1088/1748-0221/3/08/S08003",
    primaryClass   = "hep-ex",
}

@Article{SOFT-2010-01,
    author         = "{ATLAS Collaboration}",
    title          = "{The ATLAS Simulation Infrastructure}",
    journal        = "Eur. Phys. J. C",
    volume         = "70",
    year           = "2010",
    pages          = "823",
    doi            = "10.1140/epjc/s10052-010-1429-9",
    eprint         = "1005.4568",
    archivePrefix  = "arXiv",
    primaryClass   = "physics.ins-det",
}

@Article{STDM-2012-23,
    author         = "{ATLAS Collaboration}",
    title          = "{Measurement of the \(Z/\gamma^*\) boson transverse momentum distribution in \(pp\) collisions at \(\sqrt{s} = 7\,\text{TeV}\) with the ATLAS detector}",
    journal        = "JHEP",
    volume         = "09",
    year           = "2014",
    pages          = "145",
    doi            = "10.1007/JHEP09(2014)145",
    reportNumber   = "CERN-PH-EP-2014-075",
    eprint         = "1406.3660",
    archivePrefix  = "arXiv",
    primaryClass   = "hep-ex",
}

@Article{PERF-2014-03,
    author         = "{ATLAS Collaboration}",
    title          = "{Performance of pile-up mitigation techniques for jets in \(pp\) collisions at \(\sqrt{s} = 8\,\text{TeV}\) using the ATLAS detector}",
    journal        = "Eur. Phys. J. C",
    volume         = "76",
    year           = "2016",
    pages          = "581",
    doi            = "10.1140/epjc/s10052-016-4395-z",
    reportNumber   = "CERN-PH-EP-2015-206",
    eprint         = "1510.03823",
    archivePrefix  = "arXiv",
    primaryClass   = "hep-ex",
}

@Article{PERF-2015-09,
    author         = "{ATLAS Collaboration}",
    title          = "{Jet reconstruction and performance using particle flow with the ATLAS Detector}",
    journal        = "Eur. Phys. J. C",
    volume         = "77",
    year           = "2017",
    pages          = "466",
    doi            = "10.1140/epjc/s10052-017-5031-2",
    reportNumber   = "CERN-EP-2017-024",
    eprint         = "1703.10485",
    archivePrefix  = "arXiv",
    primaryClass   = "hep-ex",
}

@Article{PERF-2015-10,
    author         = "{ATLAS Collaboration}",
    title          = "{Muon reconstruction performance of the ATLAS detector in proton--proton collision data at \(\sqrt{s} = 13\,\text{TeV}\)}",
    journal        = "Eur. Phys. J. C",
    volume         = "76",
    year           = "2016",
    pages          = "292",
    doi            = "10.1140/epjc/s10052-016-4120-y",
    reportNumber   = "CERN-EP-2016-033",
    eprint         = "1603.05598",
    archivePrefix  = "arXiv",
    primaryClass   = "hep-ex",
}

@Article{TRIG-2016-01,
    author         = "{ATLAS Collaboration}",
    title          = "{Performance of the ATLAS trigger system in 2015}",
    journal        = "Eur. Phys. J. C",
    volume         = "77",
    year           = "2017",
    pages          = "317",
    doi            = "10.1140/epjc/s10052-017-4852-3",
    reportNumber   = "CERN-EP-2016-241",
    eprint         = "1611.09661",
    archivePrefix  = "arXiv",
    primaryClass   = "hep-ex",
}

@Article{DAPR-2018-01,
    author         = "{ATLAS Collaboration}",
    title          = "{ATLAS data quality operations and performance for 2015--2018 data-taking}",
    journal        = "JINST",
    volume         = "15",
    year           = "2020",
    pages          = "P04003",
    doi            = "10.1088/1748-0221/15/04/P04003",
    reportNumber   = "CERN-EP-2019-207",
    eprint         = "1911.04632",
    archivePrefix  = "arXiv",
    primaryClass   = "physics.ins-det",
}

@Article{EGAM-2018-01,
    author         = "{ATLAS Collaboration}",
    title          = "{Electron and photon performance measurements with the ATLAS detector using the 2015--2017 LHC proton--proton collision data}",
    journal        = "JINST",
    volume         = "14",
    year           = "2019",
    pages          = "P12006",
    doi            = "10.1088/1748-0221/14/12/P12006",
    reportNumber   = "CERN-EP-2019-145",
    eprint         = "1908.00005",
    archivePrefix  = "arXiv",
    primaryClass   = "hep-ex",
}

@Article{JETM-2018-05,
    author         = "{ATLAS Collaboration}",
    title          = "{Jet energy scale and resolution measured in proton--proton collisions at \(\sqrt{s} = 13\,\text{TeV}\) with the ATLAS detector}",
    journal        = "Eur. Phys. J. C",
    volume         = "81",
    year           = "2021",
    pages          = "689",
    doi            = "10.1140/epjc/s10052-021-09402-3",
    reportNumber   = "CERN-EP-2020-083",
    eprint         = "2007.02645",
    archivePrefix  = "arXiv",
    primaryClass   = "hep-ex",
}

@Article{MUON-2018-03,
    author         = "{ATLAS Collaboration}",
    title          = "{Muon reconstruction and identification efficiency in ATLAS using the full Run~2 \(pp\) collision data set at \(\sqrt{s} = 13\,\text{TeV}\)}",
    journal        = "Eur. Phys. J. C",
    volume         = "81",
    year           = "2021",
    pages          = "578",
    doi            = "10.1140/epjc/s10052-021-09233-2",
    reportNumber   = "CERN-EP-2020-199",
    eprint         = "2012.00578",
    archivePrefix  = "arXiv",
    primaryClass   = "hep-ex",
}

@Article{TRIG-2018-01,
    author         = "{ATLAS Collaboration}",
    title          = "{Performance of the ATLAS muon triggers in Run~2}",
    journal        = "JINST",
    volume         = "15",
    year           = "2020",
    pages          = "P09015",
    doi            = "10.1088/1748-0221/15/09/p09015",
    reportNumber   = "CERN-EP-2020-031",
    eprint         = "2004.13447",
    archivePrefix  = "arXiv",
    primaryClass   = "physics.ins-det",
}

@Article{TRIG-2018-05,
    author         = "{ATLAS Collaboration}",
    title          = "{Performance of electron and photon triggers in ATLAS during LHC Run~2}",
    journal        = "Eur. Phys. J. C",
    volume         = "80",
    year           = "2020",
    pages          = "47",
    doi            = "10.1140/epjc/s10052-019-7500-2",
    reportNumber   = "CERN-EP-2019-169",
    eprint         = "1909.00761",
    archivePrefix  = "arXiv",
    primaryClass   = "hep-ex",
}

@Article{FTAG-2019-07,
    author         = "{ATLAS Collaboration}",
    title          = "{ATLAS flavour-tagging algorithms for the LHC Run~2 \(pp\) collision dataset}",
    journal        = "Eur. Phys. J. C",
    volume         = "83",
    year           = "2023",
    pages          = "681",
    doi            = "10.1140/epjc/s10052-023-11699-1",
    reportNumber   = "CERN-EP-2022-226",
    eprint         = "2211.16345",
    archivePrefix  = "arXiv",
    primaryClass   = "physics.data-an",
}

@Article{DAPR-2021-01,
    author         = "{ATLAS Collaboration}",
    title          = "{Luminosity determination in \(pp\) collisions at \(\sqrt{s} = 13\,\text{TeV}\) using the ATLAS detector at the LHC}",
    journal        = "Eur. Phys. J. C",
    volume         = "83",
    year           = "2023",
    pages          = "982",
    doi            = "10.1140/epjc/s10052-023-11747-w",
    reportNumber   = "CERN-EP-2022-281",
    eprint         = "2212.09379",
    archivePrefix  = "arXiv",
    primaryClass   = "hep-ex",
}

@Article{EGAM-2021-01,
    author         = "{ATLAS Collaboration}",
    title          = "{Electron and photon efficiencies in LHC Run~2 with the ATLAS experiment}",
    year           = "2023",
    reportNumber   = "CERN-EP-2023-182",
    eprint         = "2308.13362",
    archivePrefix  = "arXiv",
    primaryClass   = "hep-ex",
}

@Booklet{ATL-SOFT-PUB-2021-001,
    author         = "{ATLAS Collaboration}",
    title          = "{The ATLAS Collaboration Software and Firmware}",
    howpublished   = "{ATL-SOFT-PUB-2021-001}",
    url            = "https://cds.cern.ch/record/2767187",
    year           = "2021",
}

@Booklet{ATL-SOFT-PUB-2023-001,
    author         = "{ATLAS Collaboration}",
    title          = "{ATLAS Computing Acknowledgements}",
    howpublished   = "{ATL-SOFT-PUB-2023-001}",
    url            = "https://cds.cern.ch/record/2869272",
    year           = "2023",
}

@Booklet{ATLAS-CONF-2018-023,
    author         = "{ATLAS Collaboration}",
    title          = "{\(E_{\text{T}}^{\text{miss}}\) performance in the ATLAS detector using 2015--2016 LHC \(pp\) collisions}",
    howpublished   = "{ATLAS-CONF-2018-023}",
    url            = "https://cds.cern.ch/record/2625233",
    year           = "2018",
}

@Booklet{ATL-PHYS-PUB-2014-021,
    author         = "{ATLAS Collaboration}",
    title          = "{ATLAS Pythia~8 tunes to \(7~\text{TeV}\) data}",
    howpublished   = "{ATL-PHYS-PUB-2014-021}",
    url            = "https://cds.cern.ch/record/1966419",
    year           = "2014",
}

@Booklet{ATL-PHYS-PUB-2016-017,
    author         = "{ATLAS Collaboration}",
    title          = "{The Pythia~8 A3 tune description of ATLAS minimum bias and inelastic measurements incorporating the Donnachie--Landshoff diffractive model}",
    howpublished   = "{ATL-PHYS-PUB-2016-017}",
    url            = "https://cds.cern.ch/record/2206965",
    year           = "2016",
}

@Booklet{ATL-PHYS-PUB-2017-007,
    author         = "{ATLAS Collaboration}",
    title          = "{Studies on top-quark Monte Carlo modelling with Sherpa and MG5\_aMC@NLO}",
    howpublished   = "{ATL-PHYS-PUB-2017-007}",
    url            = "https://cds.cern.ch/record/2261938",
    year           = "2017",
}

@Booklet{ATL-PHYS-PUB-2017-013,
    author         = "{ATLAS Collaboration}",
    title          = "{Optimisation and performance studies of the ATLAS \(b\)-tagging algorithms for the 2017-18 LHC run}",
    howpublished   = "{ATL-PHYS-PUB-2017-013}",
    url            = "https://cds.cern.ch/record/2273281",
    year           = "2017",
}

@article{LUCID2paper,
	title = 				{The new {LUCID}-2 detector for luminosity measurement and monitoring in {ATLAS}},
	author = 			{Avoni, G. and Bruschi, M. and Cabras, G. and Caforio, D. and Dehghanian, N. and Floderus, A. and Giacobbe, B. and Giannuzzi, F. and Giorgi, F. and Grafstr\"{o}m, P. and Hedberg, V. and Lasagni Manghi, F. and Meneghini, S. and Pinfold, J. and Richards, E. and Sbarra, C. and Semprini Cesari, N. and Sbrizzi, A. and Soluk, R. and Ucchielli, G. and Valentinetti, S. and Viazlo, O. and Villa, M. and Vittori, C. and Vuillermet, R. and Zoccoli, A.},
	journal = 				{JINST},
	publisher = 		{{IOP} Publishing},
	year = 					{2018},
	month =				{7},
	volume = 			{13},
	number = 			{07},
	pages = 			{P07017},
	doi = 					{10.1088/1748-0221/13/07/P07017},
	%url = 					{https://doi.org/10.1088/1748-0221/13/07/p07017},
	%eprint = 				{2633501},
	%eprinttype = 		{cds},
	%eprintclass = 		{2633501},
}

@article{fastJetPaper,
	title =				{{FastJet} user manual},
	author =		{Cacciari, M. and Salam, G.P. and Soyez, G.},
	journal =			{Eur. Phys. J. C},
	publisher =		{Springer Science and Business Media LLC},
	year =				{2012},
	month =				{3},
	volume =		{72},
	number =		{3},
	pages = 			{1896},
	%ISSN =				{1434-6052},
	%url =				{http://dx.doi.org/10.1140/epjc/s10052-012-1896-2},
	doi =				{10.1140/epjc/s10052-012-1896-2},
	reportNumber = {CERN-PH-TH/2011-297},
	eprint =              {1111.6097},
	archivePrefix  =  {arXiv},
	primaryClass   = {hep-ph},
	SLACcitation   = {%%CITATION = ARXIV:1111.6097;%%}
}

@Booklet{Cranmer:1456844,
      author        = "Cranmer, Kyle and Lewis, George and Moneta, Lorenzo and
                       Shibata, Akira and Verkerke, Wouter",
      title         = "{HistFactory: A tool for creating statistical models for
                       use with RooFit and RooStats}",
    howpublished   = "{CERN-OPEN-2012-016}",
    url            = "https://cds.cern.ch/record/1456844",
    year           = "2023",
}

@article{JAMES1975343,
title = {Minuit - a system for function minimization and analysis of the parameter errors and correlations},
journal = {Comput. Phys. Commun.},
volume = {10},
number = {6},
pages = {343-367},
year = {1975},
%issn = {0010-4655},
doi = {10.1016/0010-4655(75)90039-9},
%url = {https://www.sciencedirect.com/science/article/pii/0010465575900399},
author = {F. James and M. Roos},
}

@article{J_ger_2013,
	doi = {10.1007/jhep04(2013)024}, 
	%url = {https://doi.org/10.1007%2Fjhep04%282013%29024},
	year = 2013,
%	month = {apr},
	%publisher = {Springer Science and Business Media {LLC}},  
	volume = {04},
        pages = {24},
        number = {4},
  	author = {Barbara Jäger and Giulia Zanderighi},
  	title = {Electroweak W+W- jj production at {NLO} in {QCD} matched with parton shower in the {POWHEG}-{BOX}},
  	journal = {JHEP}
}

@article{CMS:2022woe,
    collaboration = "CMS",
    title = "{Observation of electroweak W$^+$W$^-$ pair production in association with two jets in proton-proton collisions at $\sqrt{s}$ = 13 TeV}",
    eprint = "2205.05711",
    archivePrefix = "arXiv",
    primaryClass = "hep-ex",
    reportNumber = "CMS-SMP-21-001, CERN-EP-2022-038",
%    month = "5",
    author = "{CMS Collaboration}",
    journal = {Phys. Lett. B},
    year = {2023},
    volume = {841},
    pages = {137495},
%    issn = {0370-2693},
    doi = {10.1016/j.physletb.2022.137495},
}

@article{PhysRevD.16.1519,
  title = {Weak interactions at very high energies: The role of the Higgs-boson mass},
  author = {Lee, Benjamin W. and Quigg, C. and Thacker, H. B.},
  journal = {Phys. Rev. D},
  volume = {16},
  %issue = {5},
  pages = {1519--1531},
  numpages = {0},
  year = {1977},
  month = {9},
  publisher = {American Physical Society},
  doi = {10.1103/PhysRevD.16.1519},
  %url = {https://link.aps.org/doi/10.1103/PhysRevD.16.1519}
}

@article{PhysRevD.74.073005,
  title = {$pp\ensuremath{\rightarrow}jj{e}^{\ifmmode\pm\else\textpm\fi{}}{\ensuremath{\mu}}^{\ifmmode\pm\else\textpm\fi{}}\ensuremath{\nu}\ensuremath{\nu}$ and $jj{e}^{\ifmmode\pm\else\textpm\fi{}}{\ensuremath{\mu}}^{\ensuremath{\mp}}\ensuremath{\nu}\ensuremath{\nu}$ at $\mathcal{O}({\ensuremath{\alpha}}_{\mathrm{em}}^{6})$ and $\mathcal{O}({\ensuremath{\alpha}}_{\mathrm{em}}^{4}{\ensuremath{\alpha}}_{s}^{2})$ for the study of the quartic electroweak gauge boson vertex at CERN LHC},
  author = {\'Eboli, O. J. P. and Gonzalez-Garcia, M. C. and Mizukoshi, J. K.},
  journal = {Phys. Rev. D},
  volume = {74},
  %issue = {7},
  pages = {073005},
  numpages = {14},
  year = {2006},
  month = {10},
  publisher = {American Physical Society},
  doi = {10.1103/PhysRevD.74.073005},
  %url = {https://link.aps.org/doi/10.1103/PhysRevD.74.073005}
}
 
\clearpage
% ATLAS Collaboration author list
% Reference date of STDM-2022-06 is 2023-06-23
% Author list last updated on date 26-AUG-24
% Data extracted on 26-Aug-2024 for paper reference STDM-2022-06
% at 2:35pm
 
\begin{flushleft}
\hypersetup{urlcolor=black}
{\Large The ATLAS Collaboration}

\bigskip

\AtlasOrcid[0000-0002-6665-4934]{G.~Aad}$^\textrm{\scriptsize 102}$,
\AtlasOrcid[0000-0002-5888-2734]{B.~Abbott}$^\textrm{\scriptsize 120}$,
\AtlasOrcid[0000-0002-1002-1652]{K.~Abeling}$^\textrm{\scriptsize 55}$,
\AtlasOrcid[0000-0001-5763-2760]{N.J.~Abicht}$^\textrm{\scriptsize 49}$,
\AtlasOrcid[0000-0002-8496-9294]{S.H.~Abidi}$^\textrm{\scriptsize 29}$,
\AtlasOrcid[0000-0002-9987-2292]{A.~Aboulhorma}$^\textrm{\scriptsize 35e}$,
\AtlasOrcid[0000-0001-5329-6640]{H.~Abramowicz}$^\textrm{\scriptsize 151}$,
\AtlasOrcid[0000-0002-1599-2896]{H.~Abreu}$^\textrm{\scriptsize 150}$,
\AtlasOrcid[0000-0003-0403-3697]{Y.~Abulaiti}$^\textrm{\scriptsize 117}$,
\AtlasOrcid[0000-0002-8588-9157]{B.S.~Acharya}$^\textrm{\scriptsize 69a,69b,m}$
% ...
\bigskip
\\

% ...
$^{29}$Physics Department, Brookhaven National Laboratory, Upton NY; United States of America.\\
% ...
$^{35}$$^{(a)}$Facult\'e des Sciences Ain Chock, Universit\'e Hassan II de Casablanca;$^{(b)}$Facult\'{e} des Sciences, Universit\'{e} Ibn-Tofail, K\'{e}nitra;$^{(c)}$Facult\'e des Sciences Semlalia, Universit\'e Cadi Ayyad, LPHEA-Marrakech;$^{(d)}$LPMR, Facult\'e des Sciences, Universit\'e Mohamed Premier, Oujda;$^{(e)}$Facult\'e des sciences, Universit\'e Mohammed V, Rabat;$^{(f)}$Institute of Applied Physics, Mohammed VI Polytechnic University, Ben Guerir; Morocco.\\
% ...
$^{49}$Fakult\"{a}t Physik , Technische Universit{\"a}t Dortmund, Dortmund; Germany.\\
% ...
$^{55}$II. Physikalisches Institut, Georg-August-Universit\"{a}t G\"ottingen, G\"ottingen; Germany.\\
% ...
$^{69}$$^{(a)}$INFN Gruppo Collegato di Udine, Sezione di Trieste, Udine;$^{(b)}$ICTP, Trieste;$^{(c)}$Dipartimento Politecnico di Ingegneria e Architettura, Universit\`a di Udine, Udine; Italy.\\
% ...
$^{102}$CPPM, Aix-Marseille Universit\'e, CNRS/IN2P3, Marseille; France.\\
% ...
$^{117}$Department of Physics, New York University, New York NY; United States of America.\\
% ...
$^{120}$Homer L. Dodge Department of Physics and Astronomy, University of Oklahoma, Norman OK; United States of America.\\
% ...
$^{150}$Department of Physics, Technion, Israel Institute of Technology, Haifa; Israel.\\
$^{151}$Raymond and Beverly Sackler School of Physics and Astronomy, Tel Aviv University, Tel Aviv; Israel.\\
% ...
$^{a}$ Also Affiliated with an institute covered by a cooperation agreement with CERN.\\
$^{b}$ Also at An-Najah National University, Nablus; Palestine.\\
$^{c}$ Also at Borough of Manhattan Community College, City University of New York, New York NY; United States of America.\\
$^{d}$ Also at Center for High Energy Physics, Peking University; China.\\
$^{e}$ Also at Center for Interdisciplinary Research and Innovation (CIRI-AUTH), Thessaloniki; Greece.\\
$^{f}$ Also at Centro Studi e Ricerche Enrico Fermi; Italy.\\
% ...
$^{m}$ Also at Department of Physics, King's College London, London; United Kingdom.\\
% ...

\end{flushleft}

% Created with Glance <Atlas.Glance@cern.ch>

\end{document}